\documentclass[11pt]{article}
\usepackage{amsmath}
\usepackage{mathrsfs}
\usepackage{braket}
\usepackage{ascmac}
\usepackage{bm}
\usepackage[dvipdfmx]{graphics}
\usepackage[dviout]{graphicx}
\usepackage{hyperref}
\usepackage[at]{easylist}
\usepackage{here}
\usepackage{cite}
\usepackage{amssymb}
\usepackage[version=4]{mhchem}
\usepackage{physics}
\usepackage{authblk}
\usepackage{indentfirst}
\usepackage[subrefformat=parens]{subcaption}
\usepackage[top=30truemm,bottom=30truemm,left=18truemm,right=18truemm]{geometry}
\usepackage{color}
\usepackage{ulem}

\usepackage{pagecolor}
\definecolor{darkpastelgreen}{rgb}{0.01, 0.75, 0.24}

\definecolor{DarkBlue}{rgb}{0,0,0.9}

\definecolor{DarkRed}{rgb}{0.7,0,0}

\makeatletter
\@addtoreset{equation}{section}

\makeatother

\title{Revisiting spins of primordial black holes in a matter-dominated era based on peak theory}
\date{ }
\author{Daiki Saito$^{1}$\footnote{saito.daiki.g3@s.mail.nagoya-u.ac.jp}}
\author{Tomohiro Harada$^{2}$\footnote{harada@rikkyo.ac.jp}}
\author{Yasutaka Koga$^{3}$\footnote{yasutaka.koga@yukawa.kyoto-u.ac.jp}}
\author{Chul-Moon Yoo$^{1}$\footnote{yoo.chulmoon.k6@f.mail.nagoya-u.ac.jp}}

\affil{$^{1}$Division of Science, Graduate School of Science, Nagoya University, Nagoya 464-8602, Japan}
\affil{$^{2}$Department of Physics, Rikkyo University, Toshima, Tokyo 171-8501, Japan} 
\affil{$^{3}$Yukawa Institute for Theoretical Physics, Kyoto University, Kyoto 606-8502, Japan} 

\begin{document}

\maketitle

\begin{abstract}
We estimate the probability distribution for the spins of the primordial black holes (PBHs) that formed during an early matter-dominated era in the Universe.
We employ the Zel'dovich approximation and focus on the linear-order effect of cosmological perturbations which causes the tidal torque.
Assuming that the fluctuations obey Gaussian statistics, we apply the peak theory of random Gaussian variables to compute the root mean square (RMS) and the probability distribution of the non-dimensional Kerr parameter $a_{*}$ at their formation. 
The value of $a_{*}$ is evaluated through the angular momentum at the turn-around time. 
We find that the RMS $\bar{a}_{*}$ with a given amplitude of the fluctuation $\delta_{\textmd{pk}}$ decreases as the amplitude increases.
This behavior allows us to set the threshold value of the amplitude of the fluctuation through the under-extremal condition $\bar{a}_{*}<1$.
Then we discuss the impact of spin and anisotropic collapse on the production rate of PBHs.
We find that, for $\sigma_{\textmd{H}}\leq 10^{-3}$ with $\sigma_{\rm H}$ being the square root of the variance of the fluctuation at the horizon reentry, the suppression from the spin effect is dominant, while the effect of anisotropy becomes more important for $\sigma_{\textmd{H}}>10^{-3}$. 
Since $\bar{a}_{*}$ can be written as a function of $\nu:=\delta_{\textmd{pk}}/\sigma_{\rm H}$, we can obtain the probability distribution of $\bar a_*$, $P(\bar a_*)$, through the probability distribution of $\nu$ characterized by a given power spectrum of the fluctuation. 
$P(\bar a_*)$ depends on $\sigma_{\rm H}$ and the parameter $\gamma$ that characterizes the width of the power spectrum. 
It is shown that, in the parameter regions of our interests, substantial values of PBH spins are expected in contrast to the PBH formation in a radiation-dominated universe.
For instance, with $\gamma=0.6$ and $\sigma_{\textmd{H}}=0.1$, $P(\bar a_*)$ takes a maximum at $\bar a_*\simeq0.25$.

\end{abstract}

\tableofcontents

\section{Introduction}
\label{Intro}

Primordial black holes (PBHs) are black holes (BHs) that have formed in the early Universe.
Recent attention has been drawn to PBHs due to their potential role as components of dark matter and as a source for binary black holes detected through gravitational waves~\cite{LIGOScientific:2016dsl,KAGRA:2021vkt}.
The observational constraints on their mass and spin distributions would provide valuable insights into inhomogeneities in the early universe.

Among several mechanisms, the most well-explored scenario is the formation of PBHs through the gravitational collapse of primordial density fluctuations in the radiation-dominated (RD) era of the Universe~\cite{Carr:1974nx}.
Following the end of inflation, cosmological perturbations reenter the Hubble horizon and start to evolve.
In this scenario, when the amplitude of the fluctuation exceeds a threshold, the gravitational attractive force overcomes the pressure gradient, and the perturbations grow to a non-linear regime, decoupling from the background evolution (turn-around), ultimately leading to PBH formation. 
The formation and the evaluation of PBH mass and spin in RD era have been explored in many researches~\cite{Carr:1975qj,Niemeyer:1999ak,Shibata:1999zs,Musco:2004ak,Musco:2008hv,Musco:2012au,Harada:2020pzb,DeLuca:2020bjf,Yoo:2024lhp}.

While much research has been conducted on the RD era, there is also the possible existence of an early matter-dominated (MD) epoch and the formation of PBHs during that era has recently gotten attention.
For example, at the end of inflation, the inflaton could undergo coherent oscillations, and the Universe can be approximately regarded as an MD universe.
PBH formation in the MD era was pioneered in Ref.~\cite{Khlopov:1980mg}.
While it has been anticipated that PBH formation in the MD era could be more efficient than in the RD era due to the absence of pressure gradients that halt gravitational collapse, Ref.~\cite{Khlopov:1980mg} pointed out the importance of the effect of anisotropy in gravitational collapse.
In Ref.~\cite{Harada:2016mhb}, the authors refined the discussion from Ref.~\cite{Khlopov:1980mg} by incorporating the hoop conjecture~\cite{Thorne:1972ji} and derived the production rate of PBHs considering the effect of anisotropy.
In addition, in Ref.~\cite{Harada:2017fjm},
Harada, Yoo, Kohri and Nakao (HYKN) evaluated the angular momentum generated by the tidal torque of the fluid and discussed its effect on PBH formation.
In particular, they proposed setting a threshold for density fluctuation by imposing that the spins of PBHs should be smaller than that of the extremal Kerr black hole. 
They also discussed the effect of spin and anisotropy on the production rate of PBHs and found that, for small values of $\sigma_{\textmd{H}}$, the suppression due to the spin is relevant, where  $\sigma_{\textmd{H}}^2$ represents the variance of the density fluctuation at the horizon reentry.

The discussion of angular momentum in PBH is also valuable from an observational perspective.
From the gravitational wave observations, we can measure the chirp mass, mass ratio, and effective spin of black hole binaries. Therefore, to link PBH formation scenarios with observational data~\cite{LIGOScientific:2016dsl,KAGRA:2021vkt}, a quantitative estimation of the masses and spins of PBHs is essential (See Refs.~\cite{KAGRA:2021duu,Galaudage:2021rkt} and references therein for the estimation of the effective spin of the binaries from the observations).
In this paper, we will evaluate the initial values of the PBH spins soon after those formations in the MD epoch.
In HYKN, the typical absolute value of the angular momentum of the collapsing region in MD era has been estimated by taking the ensemble average equally weighting all possible realizations. 
However, it is known that higher peaks tend to have a more spherical shape according to the peak theory of random Gaussian variables~\cite{Bardeen:1985tr,Heavens:1988}.
Since the density perturbation must exceed a threshold value for PBH formation, the correlation between the peak height and the angular momentum of the collapsing region must be taken into account for a more accurate estimation of PBH spin. 
In this paper, we improve the analyses given in HYKN taking the conditional probability of parameters characterizing the shape of the density peak with a given amplitude of the peak. 
For this purpose, we follow the methods in Refs.~\cite{White:1984uf, Catelan:1996hv}, where the angular momentum of galaxies was computed using the Zel'dovich approximation and the peak theory.
The evaluation of PBH spin using the peak theory was also discussed in some previous works for the RD era~\cite{DeLuca:2019buf,Harada:2020pzb} and the Universe dominated by a perfect fluid with the linear equation of state $p=w\rho$ ($0<w\leq1/3$)~\cite{Saito:2023fpt}.

This paper is organized as follows.
In Sec.~\ref{Pre}, we briefly review linear perturbations in the context of Newtonian gravity and introduce the Zel'dovich approximation.
In Sec.~\ref{AM}, we derive expressions for the angular momentum within a collapsing region around a peak of the density perturbation and then estimate the ensemble average of the non-dimensional spin under a specific condition of the turn-around. 
The impact of the threshold given by the under-extremal condition of the spin on the PBH abundance is discussed in Sec.~\ref{beta}.
The probability distribution function for the non-dimensional spin parameter is presented in Sec.~\ref{E_AM}, and the joint distribution for the mass and the spin parameter is discussed in Sec.~\ref{Pam}.
Sec.~\ref{Sum} is dedicated to summarizing and discussing the findings. 
Throughout this paper, we use geometrized units in which both the speed of light and Newton's gravitational constant are unity, $c=G=1$.

\section{Cosmological perturbation theory in Newtonian gravity}
\label{Pre}

In this section, we will briefly review cosmological perturbation theory within the framework of Newtonian gravity.

\subsection{Basic equations and background solution}

Basic equations are given as follows:
\begin{align}
  &\frac{\partial\rho}{\partial t}+\nabla_{\vec{r}}\cdot\qty(\rho\frac{d\vec{r}}{dt})=0, \label{eq:con} \\
  &\frac{d^2\vec{r}}{dt^2}=-\nabla_{\vec{r}}\Psi, \label{eq:Eur} \\
  &\Delta_{\vec{r}}\Psi=4\pi\rho. \label{eq:Poi}
\end{align}
The first, second, and third equations are the continuity equation, the Euler equation, and the Poisson equation, respectively.
Here, $\vec{r}$, $\nabla_{\vec{r}}$ and $\Delta_{\vec{r}}$ are the Eulerian coordinates, the derivative and Laplacian operators for $\vec{r}$, respectively.
We shall assume the background is homogeneous and isotropic, introduce the scale factor $a(t)$ and define the comoving coordinates $\vec{x}$ by $\vec{x}:=\vec{r}/a(t)$.
Then, the equations~\eqref{eq:con}, \eqref{eq:Eur} and \eqref{eq:Poi} can be rewritten as
\begin{align}
    &\dot{\rho}_{\textmd{b}}+3\frac{\dot{a}}{a}\rho_{\textmd{b}}=0, \\
    &\frac{\ddot{a}}{a}=-\frac{4\pi}{3}\rho_{\textmd{b}}, \\
    &\Delta_{\vec{r}}\Psi_{\textmd{b}}=4\pi\rho_{\textmd{b}},
\end{align}
respectively.
Here, $\rho_{\textmd{b}}$ and $\Psi_{\textmd{b}}$ denote the energy density and the gravitational potential for the background, respectively, and the dot denotes the time-derivative.
These equations can be solved to yield
\begin{align}
    &\rho_{\textmd{b}}a^3=:\eta_{0}=\textmd{Const.}, \label{eq:enecons} \\
    &\Psi_{\textmd{b}}=\frac{2\pi}{3}\rho_{\textmd{b}}r^2, \label{eq:BGpsi} \\
    &\qty(\frac{\dot{a}}{a})^2=\frac{8\pi}{3}\rho-\frac{K}{a^2}, \label{eq:FRLW1}
\end{align}
where $K$ is a constant of integration corresponding to spatial curvature. 
In the following of this paper, we shall assume that the background is flat and set $K=0$.
Combining Eqs.~\eqref{eq:FRLW1} and \eqref{eq:enecons}, we obtain
\begin{align}
    &a(t)=a_{0}t^{2/3}, \label{eq:BGa} \\
    &\rho_{\textmd{b}}(t)=\frac{1}{6\pi t^2}. \label{eq:BGrho}
\end{align}
Here, we have fixed the integration constant for $a(t)$ by $a(0)=0$.

\subsection{Eulerian perturbations and linear solutions}

Let us discuss the linear perturbation around the background solutions~\eqref{eq:BGpsi},~\eqref{eq:BGa} and ~\eqref{eq:BGrho}.
We shall define the perturbations for the gravitational potential and the energy density as $\psi:=\Psi-\Psi_{\textmd{b}}$ and $\delta:=\frac{\rho-\rho_{\textmd{b}}}{\rho_{\textmd{b}}}$, respectively.
The equations can be written as
\begin{align}
    &\frac{\partial}{\partial t}\vec{u}+\frac{\dot{a}}{a}\vec{u}+\frac{1}{a}(\vec{u}\cdot\nabla)\vec{u}=-\frac{1}{a}\nabla\psi, \label{eq:pertu0} \\
    &\frac{\partial}{\partial t}\delta+\frac{1}{a}\qty[\nabla\cdot\vec{u}+\nabla\cdot(\delta\vec{u})]=0, \label{eq:pertdel} \\
    &\Delta\psi=4\pi\rho_{\textmd{b}}a^2\delta, \label{eq:pertPoi}
\end{align}
where $\nabla$ and $\Delta$ are the derivative and Laplacian operators for $\vec x$, and we have defined the peculiar velocity $\vec{u}:=a\frac{\mathcal D\vec{x}}{dt}$ with $\mathcal D/dt:=\partial_{t}+\vec{u}\cdot \nabla /a$ being the time derivative along the motion of the fluid element.
By expanding Eqs.~\eqref{eq:pertu0} and \eqref{eq:pertdel} to linear order with respect to $\psi$, $\delta$ and $\vec{u}$, we obtain
\begin{align}
    &\frac{\partial}{\partial t}\vec{u}+\frac{\dot{a}}{a}\vec{u}=-\frac{1}{a}\nabla\psi, \label{eq:pertu} \\
    &\frac{\partial}{\partial t}\delta+\frac{1}{a}\nabla\cdot\vec{u}=0, 
\end{align}
In the Fourier space, we can solve the linear order equations as
\begin{align}    &\delta_{\vec{\bar{k}}}=A_{\vec{\bar{k}}}t^{2/3}+B_{\vec{\bar{k}}}t^{-1}, \label{eq:delL} \\
    &\vec{u}_{\vec{\bar{k}}}=a_{0}\frac{i\vec{\bar{k}}}{k^2}\qty[\frac{2}{3}A_{\vec{\bar{k}}}t^{1/3}-B_{\vec{\bar{k}}}t^{-4/3}]+\vec{C}_{\vec{\bar{k}}}t^{-2/3}, \\
    &\psi_{\vec{\bar{k}}}=-\frac{2}{3}\frac{a^2_{0}}{k^2}\qty[A_{\vec{\bar{k}}}+B_{\vec{\bar{k}}}t^{-5/3}],
\end{align}
where $A_{\vec{\bar{k}}}$ and $B_{\vec{\bar{k}}}$ are constants for integration.
$\vec{\bar{k}}$ is the comoving Eulerian wave vector and $\vec{C}_{\vec{\bar{k}}}$ satisfies $\vec{\bar{k}}\cdot\vec{C}_{\vec{\bar{k}}}=0$.
Here, we have added a bar to the Eulerian wave vector to distinguish it from that in the Lagrangian frame, which we will introduce in the next subsection.
We note that the density perturbation $\delta$ in Newtonian gravity is equivalent to that in the synchronous comoving gauge in relativistic cosmology.

\subsection{Lagrange perturbations and the Zel'dovich approximation}

In the previous subsection, we derived the linear perturbation solutions based on the Eulerian frame.
In this subsection, we discuss the linear perturbation equations and their solutions in the Lagrangian frame, which will be predominantly used in this paper
(See also Ref.~\cite{Catelan:1996hv} for the solutions of the Lagrange perturbation and their use for the tidal torque generated by the cosmological perturbation).

The Eulerian comoving coordinates $\vec{x}$ and the Lagrangian comoving coordinates $\vec{q}$ are related by 
\begin{align}
    \vec{x}=\vec{q}+\vec{D}(t, \vec{q}),
    \label{eq:EL}
\end{align}
where $\vec{D}(t,q)$ is referred to as the displacement vector, which represents the perturbation from the static configuration of the fluid.
From Eq.~\eqref{eq:pertu}, we obtain 
\begin{align}
    &\frac{\mathcal D^2}{dt^2}\vec{D}+2\frac{\dot{a}}{a}\frac{\mathcal D}{dt}\vec{D}=-\frac{1}{a^2}\nabla\psi, \label{eq:D} 
\end{align}
where we have used $\vec{u}=a\frac{\mathcal D}{dt}\vec{D}$.

Using the conservation of mass in a fixed volume, we can express the density perturbation in terms of the displacement vector:
  \begin{align}
      \rho d^3\vec{x}&=\rho_{\rm b}(1+\delta) d^3\vec{x}=\rho_{\textmd{b}} d^3\vec{q} \\
      \therefore \delta&=\frac{1}{\det(\delta_{ij}+\frac{\partial  D_{i}}{\partial  q_{j}})}-1. \label{eq:delJ}
      \end{align}
Here, we have assumed that the mean value of the energy density can be approximated to be the background value  $\rho_{\textmd{b}}$, and the value of $\rho$ is equal to $\rho_{\textmd{b}}$ in the initial time of the evolution.
Then, from Eq.~\eqref{eq:pertPoi}, we obtain
      \begin{align}
        &\Delta \psi=4\pi\rho_{\textmd{b}}a^2\qty[\frac{1}{\det(\delta_{ij}+\frac{\partial  D_{i}}{\partial  q_{j}})}-1].  \label{eq:psiL}
    \end{align}

In this paper, we will employ the Zel'dovich approximation.
In this approximation, we assume that the motions of fluid elements are determined by the gradient of the gravitational potential with respect to the Lagrangian coordinates.
This can be realized by solving the linearized equation for $\vec{D}$ to obtain the linear growth of the displacement.
Then, we shall extend the solution beyond the linear regime in terms of the Eulerian density fluctuation, $\delta(t,\vec{x})$.
Introducing the derivative operators $\Delta_{\vec{q}}$ and $\nabla_{\vec{q}}$ for the Lagrange coordinates $\vec q$ and 
linearizing Eq.~\eqref{eq:psiL}, we obtain
\begin{align}    &\Delta_{\vec{q}}\psi(\vec{q})=-4\pi\rho_{\textmd{b}}a^2\nabla_{\vec{q}}\cdot\vec{D},  \label{eq:Deqpsi}
\end{align}
which can be solved as
\begin{align}
  \vec{D}(t,\vec{q})&=-\frac{a}{4\pi\eta_0}\nabla_{\vec{q}}\psi(\vec{q})\propto t^{2/3}. \label{eq:Dpsi}
  \end{align}
Let us note that $\eta_0$ has been defined in Eq.~\eqref{eq:enecons}.
Under the Zel'dovich approximation, the density fluctuation in the Lagrangian frame is given as
\begin{align}
    \delta(t, \vec{q})=\frac{a}{4\pi\eta_0}\Delta_{\vec{q}}\psi(t,\vec{q})=-\nabla_{\vec{q}}\cdot\vec{D}(t,\vec{q}).  \label{eq:deltatq}
\end{align}

 \section{Angular momentum and root mean square of spin at the turnaround}
\label{AM}

In this section, we define the angular momentum and the mass inside a collapsing region which contains the peak of density fluctuation.

\subsection{Angular momentum and spin parameter}
\label{agmt}

In this subsection, we will derive the analytic expression for the angular momentum and the non-dimensional spin parameter.
The computations here will follow a similar approach to those in Ref.~\cite{Catelan:1996hv}, where the angular momentum of a galaxy is evaluated.

In the Eulerian frame, angular momentum in a comoving volume $V$ is given as
\begin{align}
    \vec{S}&=\int_{a^3V}d^3\vec{r}\rho\vec{r}\times\frac{d}{dt}
    \left(a\vec x\right)
    \nonumber \\ 
    &=\int_{V}a^3d^3\vec{x}\rho_{\textmd{b}}(1+\delta(t,\vec{x}))a\vec{x}\times\qty(\dot{a}\vec{x}+a\dot{\vec{x}}),
 \label{eq:AMdef}
\end{align}
where $\rho$ is energy density and $\rho_{\textmd{b}}$ denotes its background value.
In the Lagrangian frame, in the first-order of the displacement, the angular momentum can be computed as
\begin{align}
    \vec{S}&=\rho_{\textmd{b}}a^4\int_{\Gamma}d^3\vec{q}\qty(
    \vec{q}+\vec{D})\times a
    \frac{\mathcal D}{d t}
    \qty(\vec{q}+\vec{D}) \nonumber \\ 
    &\simeq\eta_{0}a^2\frac{2}{3t}\int_{\Gamma}d^3\vec{q}(\vec{q}\times\vec{D}), \label{eq:AMLag}
\end{align}
where we have used Eq.~\eqref{eq:Dpsi} and $\Gamma$ denotes the comoving volume in the Lagrange frame.
Since our goal is to estimate the PBH spin, we assume that the region of integration encloses a local density maximum $\delta_{\textmd{pk}}$ and fix the coordinates so that it is located at $\vec{q}=0$.
By using Eq.~\eqref{eq:Dpsi}, we can rewrite the components of $\vec{S}$ as
\begin{align}
    S^{i}&\simeq-\eta_{0}t\int_{\Gamma}d^3\vec{q}\qty(\vec{q}\times\nabla_{\vec{q}}\psi(\vec{q}))_{i} \nonumber \\
    &\simeq -\eta_{0}t\epsilon_{ijk}
    \qty(\int_{\Gamma}d^3\vec{q}q^{j}q^{l})\frac{\partial^2}{\partial q^{k}\partial q^{l}}\psi(0),  \nonumber \\
    S^{i}&=\eta_{0}t\epsilon_{ijk}\mathcal{D}_{jl}V^{kl}, \label{eq:AMLag2}
\end{align}
which is the equivalent expression to Eq.~(7) in Ref.~\cite{Catelan:1996hv}.
Here, we have expanded $\psi$ around the point $\vec{q}=\vec{0}$ up to the second order in $\vec{q}$ and used $\frac{\partial}{\partial q^{k}}\psi(0)=0$, by assuming that the wavelength of the fluctuation is larger than the length scale of the integrated region (See Ref.~\cite{Harada:2017fjm} for the detailed discussion).
We have defined the inertia tensor
\begin{align}
    V^{jl}&:=\int_{\Gamma}d^3\vec{q}q^{j}q^{l},
\end{align}
and the deformation tensor
\begin{align}
    \mathcal{D}_{kl}&:=\frac{\partial^2}{\partial q^{k}\partial q^{l}}\psi(0). 
\end{align}
It should be noted that, in this paper, the deformation tensor is defined as the derivative of the displacement vector. In contrast, some references, such as Refs.~\cite{Harada:2016mhb,Dalianis:2020gup}, use the same term to refer to the Lagrangian derivative of the Eulerian coordinates.
From Eq.~\eqref{eq:AMLag2}, we can see that the angular momentum grows linearly in time.
In the following of this paper, we shall take the coordinate axes of $q^{i}$ as the principal axes of $V^{jl}$ and thus $V^{jl}=0$ ($j\neq l$).
Then, we obtain
\begin{align}
  S^1&\propto \mathcal D_{23}\left(V^{33}-V^{22}\right), \\ 
  S^2&\propto \mathcal D_{31}\left(V^{11}-V^{33}\right), \\ 
  S^3&\propto \mathcal D_{12}\left(V^{22}-V^{11}\right).   
\end{align}
This relation means that the angular momentum is non-vanishing only if the lengths of the axes of the ellipsoid are different, and the deformation tensor is misaligned with the direction of the principal axes of $q^{i}$.

Hereafter, we assume that the gravitational potential $\psi$ is a random Gaussian variable. 
Then, in the case of linear approximation, all perturbative variables can be regarded as random Gaussian fields. 
Therefore, we consider the density perturbation $\delta$ as a Random Gaussian variable with the power spectrum satisfying 
  \begin{align}
    \big\langle \delta_{\vec{k}}\delta^{*}_{\vec{k}'} \big\rangle&=(2\pi)^3\delta^{3}\qty(\vec{k}-\vec{k}')\frac{2\pi^2}{k^3}\mathcal P(k), 
     \label{eq:deltaGauss}
  \end{align}
where $\vec k$ is the wave vector for the Lagrangian coordinates, and $\delta_{\vec k}$ is the Fourier transform of the 
linear density perturbation.

Let us assume that the whole overdensity region will collapse and fix the boundary of the region of integration by the condition $\delta=0$.
By expanding $\delta$ around $\vec{q}=\vec{0}$ up to the second order in $q$, we obtain
\begin{align}
  \delta&=\delta_{\rm pk}-\frac{1}{2}\sigma_2\sum_{i=1,2,3}\lambda_i (q^i)^2,  
\end{align}
where $\delta_{\textmd{pk}}:=\delta|_{\vec q=\vec 0}$, and $\lambda_{i}$ $(i=1,2,3)$ are the eigenvalues of $-\zeta_{ij}/\sigma_{2}$ with
  \begin{align}
      &\sigma^2_{j}:=\int d\ln k k^{2j}\mathcal P(k), \\
      &\zeta_{ij}:=\left.\frac{\partial^2\delta}{\partial q^{i}\partial q^{j}}\right|_{\vec{q}=\vec{0}}.  
    \end{align}
Then, the boundary of the overdense region is given by an ellipsoid: 
\begin{align}
    &\frac{q^2_{1}}{A^2_{1}}+\frac{q^2_{2}}{A^2_{2}}+\frac{q^2_{3}}{A^2_{3}}=1, \label{eq:ElipA} \\
    &A^2_{i}=2\frac{\sigma_{0}}{\sigma_{2}}\frac{\nu}{\lambda_{i}} \quad (i=1,2,3), \label{eq:A2}
  \end{align}
where we have defined $\nu:=\delta_{\textmd{pk}}/\sigma_0$.
In the following, we shall assume that $\lambda_{i}$ are ordered as $\lambda_{1}\geq\lambda_{2}\geq\lambda_{3}$.
Using Eqs.~\eqref{eq:ElipA} and \eqref{eq:A2}, we obtain
\begin{align}
    \Gamma&=\frac{4}{3}\pi r^3_{0}, \label{eq:Gamma} \\
    V^{jl}&=\int_{\Gamma}d^3\vec{q}q^{j}q^{l} \nonumber \\
    &=\frac{1}{5}\Gamma\textmd{diag}\qty(A^2_{1},A^2_{2},A^2_{3}) ,\nonumber \\
    &=:\textmd{diag}(i_{1},i_{2},i_{3}), 
    \label{eq:V}
\end{align}
where we have defined the mean radius of the overdense region as $r_{0}:= \qty(A_{1}A_{2}A_{3})^{1/3}$.

Let us define the mass of the overdensity as the product of the background energy density and the physical volume of the overdense region as
\begin{align}
    M&:=\frac{4\pi}{3}\rho_{\textmd{b}}(ar_{0})^3=\Gamma\eta_{0}. \label{eq:M}
\end{align}
Using this definition, we can introduce the non-dimensional Kerr parameter as
\begin{align}
    a_{*i}:=\frac{S_{i}}{M^2}=\frac{t\epsilon_{ijk}\mathcal{D}_{jl}V^{kl}}{\Gamma^2\eta_{0}}.
\end{align}

\subsection{Root mean square of spin}
\label{Typ}

In the previous subsection, we derived analytical expressions for the angular momentum and spin (non-dimensional Kerr parameter) generated by tidal torques within a closed region. 
In this subsection, we derive the root mean square (RMS) value of the spin based on the peak statistics of fluctuations~\cite{Bardeen:1985tr,Heavens:1988} (See App.~\ref{PT} for a brief review).

The squared spin
\begin{align}   
a_{*}^2:=a_{*}^{i}a_{*i}&=\qty(\frac{t}{\Gamma^2\eta_{0}})^2\epsilon_{ijk}\epsilon_{imn}V^{kl}V^{ns}\mathcal{D}_{jl}\mathcal{D}_{ms}
\label{eq:as2}
  \end{align}
depends on the Gaussian variables via $V^{ij}$ and $\mathcal{D}_{ij}$. 
Thus, we shall consider their correlation.
According to the peak theory, the statistical profile of the off-diagonal part of $\mathcal{D}_{ij}$ is independent of that of $(\nu, \lambda_{i})$ (See Eq.~\eqref{eq:Pw}).
From Eqs.~\eqref{eq:Gamma} and \eqref{eq:V}, we see that
$\Gamma$ and $V^{ij}$ depend only on $\lambda_{i}$.
Thus, we can take the ensemble average of the products of the off-diagonal components of the deformation tensor $\mathcal{D}_{jl}\mathcal{D}_{ms}$ and that of the other part separately.
The ensemble average of $\mathcal{D}_{ij}\mathcal{D}_{kl}$ can be calculated as (See Eq.~\eqref{eq:DDA})
\begin{align}
    \langle \mathcal{D}_{ij}\mathcal{D}_{kl} \rangle_{w_{i}}&=\frac{1-\gamma^2}{15}(4\pi\rho_{\textmd{b}}a^2\sigma_{0})^2\qty(\delta_{ik}\delta_{jl}+\delta_{il}\delta_{jk}+\delta_{ij}\delta_{kl}) \nonumber \\  &=:\Phi\qty(\delta_{ik}\delta_{jl}+\delta_{il}\delta_{jk}+\delta_{ij}\delta_{kl}), \label{eq:DD} 
      \end{align}
where the notation $\big\langle \bullet\big\rangle_{w_{i}}$ denotes the ensemble average over $w_{i}$, the normalized off-diagonal part of $\mathcal{D}_{ij}$ (see App.~\ref{PT} for the explicit definition of $w_i$).
We have also defined
\begin{align}
    \gamma&:=\frac{\sigma^2_{1}}{\sigma_{0}\sigma_{2}},
  \end{align}
which characterizes the width of the power spectrum of the gravitational potential.
Then, the average of the squared spin over $\mathcal{D}_{ij}$ can be expressed as
\begin{align}
    a_{\lambda}^2:=\big\langle a_{*}^2 \big\rangle_{w_{i}}&=\qty(\frac{t}{\eta_{0}})^2\epsilon_{ijk}\epsilon_{imn} \frac{V^{kl}V^{ns}}{\Gamma^4} \cdot\Phi
    \qty(\delta_{jm}\delta_{ls}+\delta_{js}\delta_{lm}+\delta_{jl}\delta_{ms})  \nonumber \\
    &=\qty(\frac{t}{\eta_{0}})^2\Phi\frac{3V^{jl}V^{jl}-\qty(V^{jj})^2}{\Gamma^4}  \nonumber \\
    &=2\qty(\frac{t}{\eta_{0}})^2\Phi\frac{\mu^2_{1}-3\mu_{2}}{\Gamma^4},
     \label{eq:AMS}
  \end{align}
where $\mu_{1}:=i_{1}+i_{2}+i_{3}$ and $\mu_{2}:=i_{1}i_{2}+i_{2}i_{3}+i_{3}i_{1}$.

For the further evaluation, it would be useful to normalize the spin at the horizon reentry time $t_{\textmd{H}}$ satisfying
$a(t_{\textmd{H}})r_{0}=H(t_{\textmd{H}})^{-1}$.
From this relation we find $(ar_{0})^2=3Mt_{\textmd{H}}(t/t_{\textmd{H}})^{4/3}$ and $\sigma_{0}=\sigma_{\rm H}(t/t_{\textmd{H}})^{2/3}$ with $\sigma_{\textmd{H}}:=\sigma_{0}(t_{\textmd{H}})$.
Therefore, we can express the squared spin as
\begin{align}
a_{\lambda}^2=\frac{2^2\cdot 3}{5^3}(1-\gamma^2)\sigma_{\textmd{H}}^2\qty(\frac{t}{t_{\textmd{H}}})^2Q^2 \label{eq:a},
  \end{align}
where we have defined the non-dimensional quadrupole moment
\begin{align}    Q&:=\frac{\sqrt{2\qty[\mu^2_{1}-3\mu_{2}]}}{\frac{3}{5}\Gamma r_{0}^{2}}. \label{eq:q}
\end{align}
We see that $Q$ is a function of the shape parameters $\lambda_{i}$ around the density peak and can be written as 
\begin{align}
&Q(e,p)=\sqrt{2}\frac{A(e,p)^{1/2}}{B(e,p)^{2/3}}, \quad
A(e,p):=(p(1+p))^2+3e^2(1-6p+2p^2+3e^2), \quad 
B(e,p):=(1-2p)((1+p)^2-9e^2), \label{eq:Qdef} \\ 
&x=\lambda_{1}+\lambda_{2}+\lambda_{3}, \quad e:=\frac{\lambda_{1}-\lambda_{3}}{2x}, \quad p:=\frac{\lambda_{1}-2\lambda_{2}+\lambda_{3}}{2x}. 
  \end{align}
Let us further perform averaging over $\lambda_{i}$ with the value of $\nu$ fixed as follows:
\begin{align}
  \bar{a}_{*}(\nu)
  &:=\frac{2\cdot 3^{1/2}}{5^{3/2}}\sqrt{1-\gamma^2}\sigma_{\textmd{H}}\qty(\frac{t}{t_{\textmd{H}}})\sqrt{\langle Q^{2}\rangle_{\tilde{\lambda}_{i}|\nu}},
\label{eq:AMEns}
  \end{align}
where $\big\langle Q^{2} \big\rangle_{\tilde{\lambda}_{i}|\nu}$ is the ensemble average of $Q^{2}$ over $\tilde{\lambda}_{i}:=(x, e, p)$ with fixed $\nu$,
\begin{align}
\left\langle Q^{2}\right\rangle_{\tilde{\lambda}_{i}|\nu}&:=\int dxdedp Q^{2}(e,p) P(x,e,p|\nu), \label{eq:QInt}
\end{align}
with $P(x,e,p|\nu)$ being the conditional peak probability for $\tilde{\lambda}_{i}$ with fixed $\nu$ (See App.~\ref{PT} for the specific expression).

\begin{figure}[htbp]
  \begin{center}
   \includegraphics[clip,width=10cm]{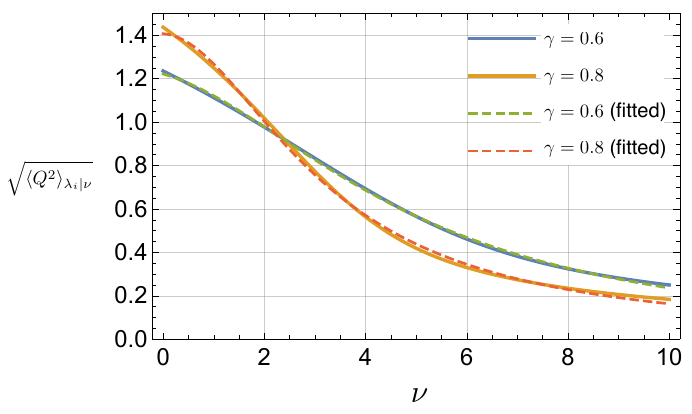} 
   \caption{The values of $\sqrt{\big\langle Q^{2} \big\rangle_{\tilde{\lambda}_{i}|\nu}}$ as functions of $\nu$.
   The solid curves denote the results of the numerical integration~\eqref{eq:QInt} for $\gamma=0.6$ (blue) and $0.8$ (yellow).
   Fitted functions $(a\nu+1)/(b\nu^2+c\nu+d)$ are also depicted as dashed curves.
   The green and red dashed curves correspond to $\gamma=0.6$ and $0.8$ cases, respectively.}
   \label{fig:Q}
\end{center}
 \end{figure}

The $\nu$ dependence of $\big\langle Q^{2} \big\rangle_{\tilde{\lambda}_{i}|\nu}$ is shown in Fig.~\ref{fig:Q}.
The blue and yellow curves denote the cases with $\gamma=0.6$ and $0.8$, respectively.
We can see that, $\big\langle Q^{2} \big\rangle_{\tilde{\lambda}_{i}|\nu}$ decreases monotonically with $\nu$, and the value at $\nu=0$ gets larger for a larger value of $\gamma$.
In the figure, we have also plotted the fitted lines $\sqrt{\big\langle Q^{2} \big\rangle_{\tilde{\lambda}_{i}|\nu}}=(a\nu+1)/(b\nu^2+c\nu+d)$ as the dashed curves.
Fitted parameters are determined as $(a,b,c,d)=(-0.0051,0.028,0.041,0.82)$ for $\gamma=0.6$ and $(a,b,c,d)=(0.014,0.047,0.042,0.71)$ for $\gamma=0.8$, respectively.
From that, we find that $\sqrt{\big\langle Q^{2} \big\rangle_{\tilde{\lambda}_{i}|\nu}}$ can be approximated as a linear function of $\nu$ for $\nu\ll1$, whereas $\sqrt{\big\langle Q^{2} \big\rangle_{\tilde{\lambda}_{i}|\nu}}\sim1/\nu$ for $\nu\gg1$.
That behavior in the large $\nu$ regime can be analytically evaluated as follows: according to the peak theory~\cite{Bardeen:1985tr,Heavens:1988}, for $\nu\gg1$, we have 
\begin{align}
\lambda_{i}\simeq\frac{\gamma\nu}{3}(1+\epsilon_{i}), \quad \epsilon_{i}=O\qty(\frac{1}{\gamma\nu}). 
\end{align}Substituting this into Eqs.~\eqref{eq:Qdef} and \eqref{eq:QInt}, we observe that
\begin{align}
&Q=O\qty(\frac{1}{\gamma\nu}), 
\\
\therefore \quad 
&\sqrt{\big\langle Q^{2} \big\rangle_{\tilde{\lambda}_{i}|\nu}}=O\qty(\frac{1}{\gamma\nu}), \label{eq:QLnu}
\end{align}
at the leading order in $1/\gamma\nu$.

\subsection{Root mean square of spin at the turnaround}
\label{vata}

As we have seen in the previous subsection, the value of $\bar a_*$ also depends on time, and thus we need to specify the evaluation time.
In this paper, we shall assume that the duration of the early MD era is longer than the time interval from the horizon reentry to the 
shell-crossing time, approximately corresponding to the PBH formation time, and we will evaluate the spin at the turn-around time $t_{\textmd{ta}}$.
The turn-around time $t_{\textmd{ta}}$ is a time when the collapsing region decouples from the background expansion of the universe, and the mass and the total angular momentum remain conserved after that.
While there is no definitive criterion for identifying the turn-around time for growing non-spherical overdensities, we shall define it by referring to the results for spherical collapse based on the relativistic cosmological dust model (See App.~\ref{TA} for the derivation) as \eqref{eq:tta}
  \begin{align}
  \frac{t_{\textmd{ta}}}{t_{\textmd{H}}}=\frac{3\pi}{4}\qty(\frac{5}{3}\nu\sigma_{\textmd{H}})^{-3/2}. \label{eq:tH} 
\end{align}
We shall note that we have assumed that $\nu\sigma_{\textmd{H}}\ll1$ to derive this equation.

Then we obtain
\begin{align}
    \left.a_{\lambda}^{2}\right|_{t=t_{\textmd{ta}}}&=
    \frac{2^2 3^4}{5^6}\qty(\frac{3\pi}{4})^2(1-\gamma^2)\sigma_{\textmd{H}}^{-1}Q^{2}\nu^{-3},
     \label{eq:ALnst} 
  \end{align}
and thus
\begin{align}
    \bar{a}_{*}|_{t=t_{\textmd{ta}}}&=
    \frac{2\cdot 3^2}{5^3}
    \frac{3\pi}{4}\sqrt{1-\gamma^2}\sigma_{\textmd{H}}^{-1/2}\sqrt{\langle Q^{2}\rangle_{\lambda_{i}|\nu}}\nu^{-3/2},
     \label{eq:AMEnst} 
  \end{align}
where we have used the relation \eqref{eq:tH}.
Hereafter, we omit the subscript $t=t_{\textmd{ta}}$
unless it is needed. 
It is worth noting that the further $\nu$ dependence of $\nu^{-3/2}$ appeared from the $\nu$ dependence of the turn-around time 
$t_{\rm ta}$. 
We also observe that $\bar{a}_{*}$ includes the factor $\sqrt{1-\gamma^2}$, reflecting the statistical properties of the deformation tensor.
Since $\gamma$ characterizes the width of the power spectrum, this implies that the value of the spin $\bar{a}_{*}$ is smaller for narrower power spectra.

\begin{figure}[htbp]
  \begin{center}
   \includegraphics[clip,width=14cm]{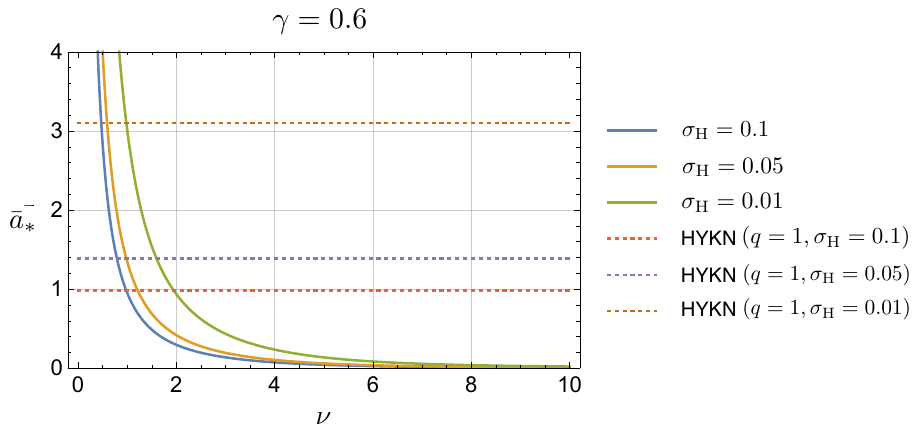} 
   \caption{
   The plot of $\bar{a}_{*}$ for $\gamma=0.6$.
   The blue, orange, and green curves denote the cases with $\sigma_{\textmd{H}}=0.1$, $0.05$ and $0.01$, respectively.
   The red, purple, and yellow dotted lines are the values of the spin parameter evaluated using Eq.~\eqref{eq:asharadatm} with $q=1$ and $\sigma_{\textmd{H}}=0.1$, $0.05$ and $0.01$, respectively.}
   \label{fig:a}
\end{center}
 \end{figure}

The $\nu$ dependence of the spin $\bar{a}_{*}$ is depicted in Fig.~\ref{fig:a}. 
We can observe that $\bar{a}_{*}$ decreases monotonically with $\nu$. 
This result reflects the fact that the non-spherical effect around the density peak is more crucial for smaller $\nu$. 
We find that, with fixed $\nu$, a smaller value of $\sigma_{\textmd{H}}$ gives a larger value of $\bar{a}_{*}$.
This can be understood as follows: the smaller $\sigma_{\textmd{H}}$ is, the longer time is needed for the fluctuation from the horizon reentry to the turn-around as is seen from Eq.~\eqref{eq:tH}, and thus the larger the angular momentum grows due to the tidal torque. 
We also find that $\bar{a}_{*}\geq 1$ for small $\nu$.
For example, with $\gamma=0.6$, $\bar{a}_{*}\geq 1$ for $\nu\leq1.0$ and $\nu\leq1.9$ with $\sigma_{H}=0.1$ and $0.01$, respectively.
In these regions of $\nu$, we suppose that the centrifugal force of the matter is so significant that PBHs cannot form.
Instead, rapidly rotating minihalos would likely develop.

At the end of this subsection, let us compare our result with that in the previous work by Harada, Yoo, Kohri, and Nakao (HYKN)~\cite{Harada:2017fjm}. 
In HYKN, the RMS of the spin parameter has been evaluated as 
\begin{equation}
  \langle a_{*}^2\rangle^{1/2}=\frac{2}{5}\sqrt{\frac{3}{5}}q\sigma_H\left(\frac{t}{t_{\rm H}}\right). 
  \label{eq:asharada}
\end{equation}
Here, although the definition of $q$ is originally the same as $Q$ in this paper, 
it has been treated as a constant parameter in HYKN.
Since the peak statistics is not taken into account in HYKN, no $\gamma$ dependence exists in the expression 
\eqref{eq:asharada}. 
There is another crucial difference in the estimation of the turnaround time between our estimation and that in HYKN.
In HYKN,
the spin parameter is estimated at the time 
\begin{equation}
t=\langle t_{\rm m}\rangle =t_{\rm H} \sigma_{\rm H}^{-3/2},
\label{eq:tm}
\end{equation} 
when $\langle\delta^2\rangle^{1/2}=1$.
Then, the RMS was computed as 
\begin{equation}
  \langle a_{*}^2\rangle^{1/2}|_{t=\langle t_{\rm m}\rangle}=\frac{2}{5}\sqrt{\frac{3}{5}}q\sigma_H^{-1/2}. 
  \label{eq:asharadatm}
\end{equation}
That is to say, in HYKN, all possible configurations have been taken into account with the same weighting, and the $\nu$-dependence of the spin parameter is simply neglected.
The values of the spin parameter $\langle a_{*}^2\rangle^{1/2}|_{t=\langle t_{\rm m}\rangle}$ with $q=1$ are depicted in Fig.~\ref{fig:a} as dotted lines.
From the figure, we can see the significant difference in the $\nu$ (in)dependence of the spin value, while its $\sigma_{\textmd{H}}$ dependence is the same for both cases which is obvious from Eqs.~\eqref{eq:AMEns} and \eqref{eq:asharadatm}.

\section{Threshold and PBH production rate}
\label{beta} 

From the expression \eqref{eq:AMEnst}, we may introduce the threshold value $\nu_{\textmd{th}}$ for the PBH formation based on the following condition
  \begin{align}
    &\bar{a}_{*}(\nu)<\bar{a}_{*}(\nu_{\textmd{th}})=1. \label{eq:nuth}
    \end{align}
    
  \begin{figure}[htbp]
      \begin{center}
       \includegraphics[clip,width= 16cm]{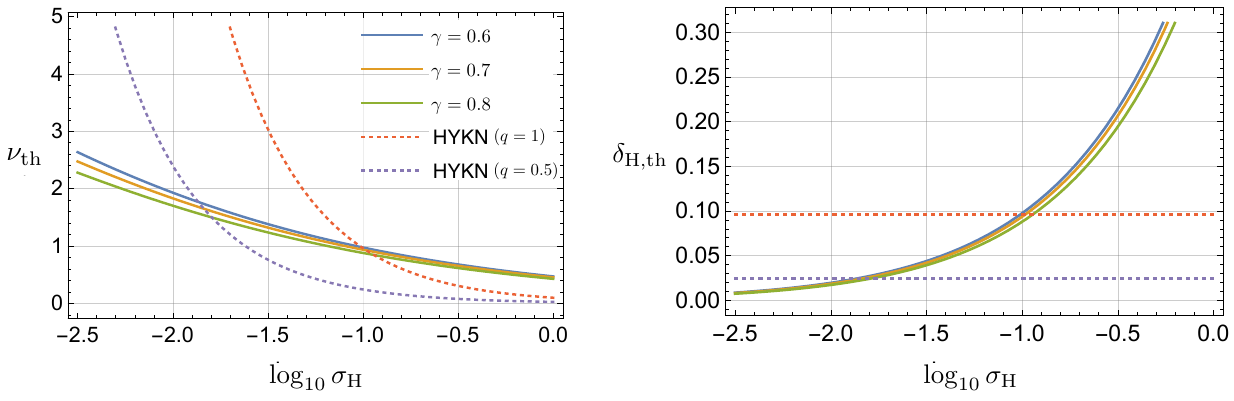} 
       \caption{The threshold of the amplitude of the density fluctuation $\nu_{\textmd{th}}$ (left) and $\delta_{\textmd{H,th}}$ (right) for the PBH formation.
       The blue, orange, and green lines denote the case with $\gamma=0.6, 0.7, 0.8$, respectively.
       The red and purple dotted curves denote the threshold value determined by  Eq.~\eqref{eq:dthHYKN} with $q=1$ and $q=0.5$, respectively.}
       \label{fig:nuth}
    \end{center}
     \end{figure}
In the left panel of Fig.~\ref{fig:nuth}, the $\sigma_{\textmd{H}}$ dependence of the threshold of the normalized amplitude $\nu_{\textmd{th}}$ is shown.
We observe that a larger $\sigma_{\textmd{H}}$ gives a smaller $\nu_{\textmd{th}}$, which can be understood from the fact that $\bar{a}_{*}\propto\sigma_{\textmd{H}}^{-1/2}$ and that $\bar{a}_{*}$ decreases with $\nu$.
We can also see the $\gamma$ dependence of the threshold value through the factor $\sqrt{1-\gamma^2}$ in Eq.~\eqref{eq:AMEnst}. 
For the monochromatic spectrum case ($\gamma=1$), angular momentum cannot be generated for the following reason.
From \eqref{eq:Pw}, we see that $P_w(\vec w) \rightarrow \delta_{D}(\vec w)$ as $\gamma \rightarrow 1$ where $\delta_{D}$ denotes the Dirac's delta function.
This implies that only the configuration with $\vec{w}=\vec{0}$ is allowed, resulting in no angular momentum being generated from the first-order effect that we are considering in this paper.
In more general, we can see that, for a fixed value of $\sigma_{\textmd{H}}$, a larger value of $\gamma$, namely, a narrower power spectrum, gives a smaller threshold. 
The threshold in terms of the density fluctuation at the horizon reentry, $\delta_{\textmd{H,th}}$, is depicted in the right panel of Fig.~\ref{fig:nuth}.
We can observe that $\delta_{\textmd{H,th}}$ is not constant and grows monotonically with $\sigma_{\textmd{H}}$.
This differs from the threshold for the spherical PBH formation against the pressure gradient in an RD universe~\cite{Carr:1974nx,Shibata:1999zs,Musco:2004ak,Musco:2008hv,Musco:2012au,Harada:2015yda,Escriva:2019phb,Escriva:2020tak}. 

Here, let us clarify the difference between our threshold estimation and that in HYKN.
In HYKN, inspired by Eq.~\eqref{eq:asharadatm}, the following equation was hypothetically assumed:
\begin{equation}
  \bar a_*^{\rm HYKN}=\frac{2}{5}\sqrt{\frac{3}{5}}q \delta_{\rm H}^{-1/2}.
  \label{eq:ashykn}
\end{equation}
Then the threshold value was estimated as 
\begin{equation}
  \delta_{\rm H, th}^{\rm HYKN}=\frac{3\cdot 2^2}{5^3}q^2.
  \label{eq:dthHYKN}
\end{equation}
We plot the value of $\delta_{\rm H, th}^{\rm HYKN}$ and $\nu_{\rm H, th}^{\rm HYKN}:=\delta_{\rm H, th}^{\rm HYKN}/\sigma_{\rm H}$ with $q=1$ and $0.5$ in Fig.~\ref{fig:nuth} as dotted curves.
Comparing $\delta_{\rm H, th}^{\rm HYKN}$ with our result shown in Fig.~\ref{fig:nuth}, one can find a crucial difference in the $\sigma_{\rm H}$ dependence of the threshold value, which will significantly affect the abundance estimation.

By using the threshold value, let us consider the impact on the PBH abundance. 
We basically follow the same procedure adopted in HYKN incorporating our estimation of the threshold. 
In HYKN the production rate of the PBHs was calculated by
\begin{align}
    \beta_{0}\simeq\int_{0}^{\infty}d\tilde{\alpha}\int_{-\infty}^{\tilde{\alpha}}d\tilde{\beta}\int_{-\infty}^{\tilde{\beta}}d\tilde{\gamma}\theta[\delta_{\textmd{H}}(\tilde{\alpha}, \tilde{\beta}, \tilde{\gamma})-\delta_{\textmd{H,th}}]\theta[1-h(\tilde{\alpha}, \tilde{\beta}, \tilde{\gamma})]w(\tilde{\alpha}, \tilde{\beta}, \tilde{\gamma}), \label{eq:beta}
  \end{align}
where $\tilde{\alpha}, \tilde{\beta}$ and $\tilde{\gamma}$ are the eigenvalues of $-\frac{\partial D_{i}}{\partial q_{j}}$ at the horizon reentry which satisfies $\tilde{\alpha}\geq\tilde{\beta}\geq\tilde{\gamma}$,
and $w$ and $h$ are defined by~\cite{Doroshkevich1970}
\begin{align}
    w\qty(\tilde{\alpha}, \tilde{\beta}, \tilde{\gamma}):=-\frac{27}{8\sqrt{5}\pi(\sigma_{\textmd{H}}/\sqrt{5})^6}\exp\qty[-\frac{3}{5(\sigma_{\textmd{H}}/\sqrt{5})^2}\qty{\tilde{\alpha}^2+\tilde{\beta}^2+\tilde{\gamma}^2-\frac{1}{2}(\tilde{\alpha}\tilde{\beta}+\tilde{\beta}\tilde{\gamma}+\tilde{\gamma}\tilde{\alpha})}] \nonumber \\
    \cdot(\tilde{\alpha}-\tilde{\beta})(\tilde{\beta}-\tilde{\gamma})(\tilde{\gamma}-\tilde{\alpha})d\tilde{\alpha} d\tilde{\beta} d\tilde{\gamma},
  \end{align}
and
  \begin{align}
    h\qty(\tilde{\alpha}, \tilde{\beta}, \tilde{\gamma}):=\frac{2}{\pi}\frac{\tilde{\alpha}-\tilde{\gamma}}{\tilde{\alpha}^2}E\qty(\sqrt{1-\qty(\frac{\tilde{\alpha}-\tilde{\beta}}{\tilde{\alpha}-\tilde{\gamma}})^2}),
  \end{align}
respectively.
Here, $E$ is the complete elliptic integral of the second kind. 
We can see from Eq.~\eqref{eq:delJ} that density fluctuation at the horizon reentry can be written as $\delta_{\textmd{H}}\simeq\tilde{\alpha}+\tilde{\beta}+\tilde{\gamma}$.
Besides the first step function to introduce the threshold due to the angular momentum, we have introduced the second step function for $h$ to impose the condition $h(\tilde{\alpha}, \tilde{\beta}, \tilde{\gamma})\leq 1$, which is the condition for the ellipticity introduced in Ref.~\cite{Harada:2016mhb} based on the hoop conjecture~\cite{Thorne:1972ji}
\footnote{
The equivalence of the computation in this paragraph and that in Ref.~\cite{Harada:2016mhb} can be confirmed as follows.
The set of eigenvalues $(\tilde{\alpha}, \tilde{\beta}, \tilde{\gamma})$ is equal to $\qty(\frac{b}{a}(t_{\rm H})\alpha, \frac{b}{a}(t_{\rm H})\beta, \frac{b}{a}(t_{\rm H})\gamma)$ in Refs.~\cite{Harada:2016mhb,Dalianis:2020gup}.
In Refs.~\cite{Harada:2016mhb,Dalianis:2020gup}, $\qty(\alpha, \beta, \gamma)$ were defined as eigenvalues of $-\frac{\partial p_{i}}{\partial q_{k}}$ with
the function $p_{i}(\vec{q})$ being introduced by the relation
\begin{align}
   r_{i}=a(t)q_{i}+b(t)p_{i}(\vec{q}).
  \end{align}
In Ref.~\cite{Harada:2016mhb}, $b(t)$ was normalized so that $b(t_{\rm H})=a(t_{\rm H})$, and then the relation
\begin{align}
   \qty(\tilde{\alpha}, \tilde{\beta}, \tilde{\gamma})=\qty(\alpha, \beta, \gamma)
  \end{align}
holds.
Thus, under this normalization, the condition corresponds to the hoop conjecture can be written as $h\qty(\tilde{\alpha}, \tilde{\beta}, \tilde{\gamma}) \leq 1$, which is an equivalent condition to Eq.~(26) of Ref.~\cite{Harada:2016mhb}.
}. 

\begin{figure}[htbp]
    \begin{center}
      \includegraphics[clip,width=18cm]{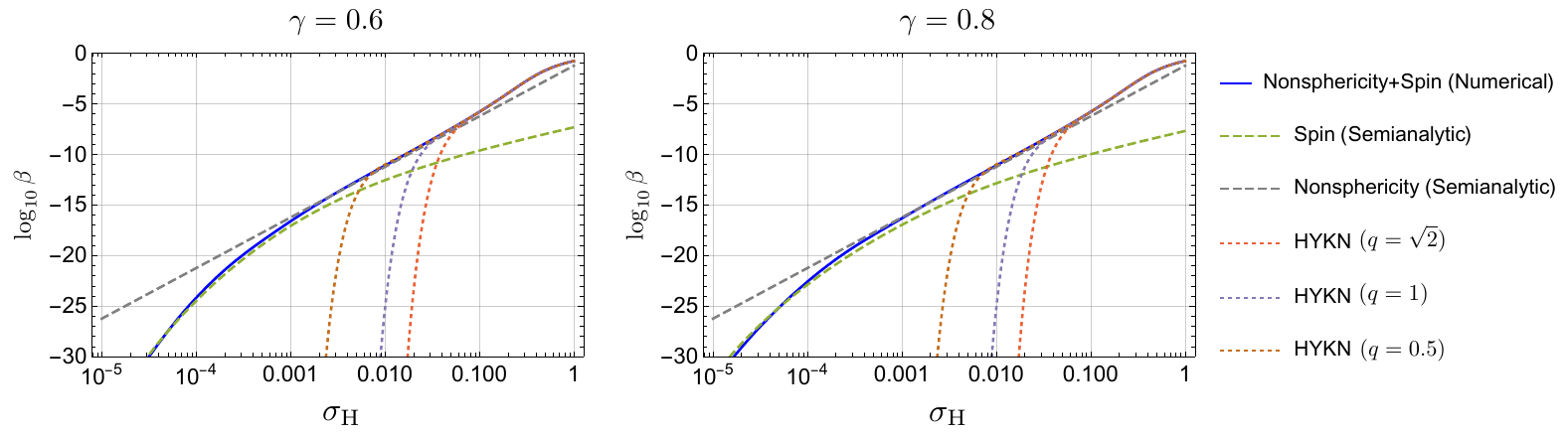}
      \caption{Production rate for PBHs with $\gamma=0.6$ and $\gamma=0.8$.
     The blue curve, the green dashed curve line, and the gray dashed curve denote the production rate in which both the effects of the spin and the non-spherical collapse are considered~\eqref{eq:beta}, the semi-analytic formula~\eqref{eq:betaana}, the rate in which only the non-spherical effect is considered~\eqref{eq:Pai}, respectively.
     The suppression of the rate due to the effect of spin can be seen for $\sigma_{\textmd{H}}\lesssim 10^{-3}$ and $\sigma_{\textmd{H}}\lesssim 4\times 10^{-4}$ for $\gamma=0.6$ and $0.8$, respectively.
     The dotted curves denote the production rates for PBH \eqref{eq:beta} with $\delta_{\rm H, th}^{\rm HYKN}$ being substituted into $\delta_{\textmd{th}}$ with $q=\sqrt{2}$ (red), $1$ (purple) and $0.5$ (yellow), respectively.}
     \label{fig:beta06}
  \end{center}
   \end{figure}

In Fig.~\ref{fig:beta06}, the production rate $\beta_{0}$ with $\gamma=0.6$ and $\gamma=0.8$ are depicited.
Here, we have substituted the obtained threshold by the condition~\eqref{eq:nuth} into $\delta_{\textmd{H,th}}$.
The blue and the green dashed lines denote the result of the numerical integration and its semianalytic formula for $\sigma_{\textmd{H}}\ll 1$~\cite{Harada:2017fjm},
\begin{align}
    \beta_{0}\simeq \frac{5\sqrt{5}\pi^4}{6^9}\bar{E}^{-5}\frac{\delta^9_{\textmd{H,th}}}{\sigma^4_{\textmd{H}}}\exp\qty(-\frac{\delta^2_{\textmd{H,th}}}{2\sigma^2_{\textmd{H}}}),
    \label{eq:betaana}
    \end{align}
respectively.
Here, $\bar{E}\simeq1.182$ and we have neglected the condition of the ellipticity in calculating Eq.~\eqref{eq:betaana}.
In Fig.~\ref{fig:beta06}, the production rate without imposing the threshold from the spin~\cite{Harada:2016mhb}
\begin{align}
    P_{\textmd{ai}}\simeq\int_{0}^{\infty}d\tilde{\alpha}\int_{-\infty}^{\tilde{\alpha}}d\tilde{\beta}\int_{-\infty}^{\tilde{\beta}}d\tilde{\gamma}\theta[1-h(\tilde{\alpha}, \tilde{\beta}, \tilde{\gamma})]w(\tilde{\alpha}, \tilde{\beta}, \tilde{\gamma})\simeq0.05556\sigma^5_{\textmd{H}} \label{eq:Pai}
    \end{align}
is also plotted by the gray dashed curve.

We can see that with $\gamma=0.6$ for $\sigma_{\textmd{H}}\gtrsim 10^{-3}$, the production rate can be fitted well by the gray dashed line, the rate considering only anisotropy \eqref{eq:Pai}.
On the other hand, for $\sigma_{\textmd{H}}\lesssim 10^{-3}$, the rate is strongly suppressed compared with $P_{\textmd{ai}}$, which implies that the effect of the threshold from the spin dominates. 
In the case with $\gamma=0.8$, the effect of spin begins to become significant at a slightly smaller value of $\sigma_{\textmd{H}}$ ($\sigma_{\textmd{H}}\lesssim 4\times 10^{-4}$). 

Let us clarify how the difference between $\delta_{\rm H,th}$ and $\delta_{\rm H,th}^{\rm HYKN}$ affects the PBH abundance estimation. 
As is shown in Fig.~\ref{fig:beta06}, for relatively small $\sigma_{\rm H}$, the suppression of the PBH abundance due to the spin effect 
is much weaker
in our case compared to that in HYKN, which are depicted as the dotted curves in the figure. 
This difference originates from the $\sigma_{\rm H}$ dependence of $\delta_{\rm H,th}$ explicitly shown in Fig.~\ref{fig:nuth}, 
which is induced by the difference between the expressions \eqref{eq:AMEnst} and \eqref{eq:asharadatm}. 
It would be worth noting that we have employed the peak statistics in Eq.~\eqref{eq:AMEns} and 
properly taken into account the amplitude dependence of the turn-around time in Eq.~\eqref{eq:tH}. 
As a result, the ensemble average of $Q^2$ obeys $\sqrt{\big\langle Q^{2} \big\rangle_{\tilde{\lambda}_{i}|\nu}}\propto 1/\nu$ for $\nu\gg1$ (See Eq.~\eqref{eq:QLnu}), and thus we can roughly estimate that
\begin{align}
\bar{a}_{*}(t_{\textmd{ta}})\simeq C_{a} \frac{\sqrt{1-\gamma^2}}{\gamma} \sigma_{\textmd{H}}^{-1/2}\nu^{-5/2}, \quad \therefore \nu_{\textmd{th}} \simeq C_{a}^{2/5} \qty(\frac{1-\gamma^2}{\gamma^2})^{1/5}\sigma_{\textmd{H}}^{-1/5}, \label{eq:nuthapp}
\end{align}
in that regime.
Here, $C_{a}$ is a constant.
Therefore, by substituting Eq.~\eqref{eq:nuthapp} into Eq.~\eqref{eq:betaana}, we obtain
\begin{align}
    \beta_{0}\sim\exp\qty[-\frac{C_{a}^{4/5}}{2}\qty(\frac{1-\gamma^2}{\gamma^2})^{2/5}\sigma_{\textmd{H}}^{-2/5}].
    \label{eq:betasup}
    \end{align}
On the other hand, for Eqs.~\eqref{eq:asharada} and \eqref{eq:ashykn}, which are taken from HYKN, the amplitude dependence of the turn-around time and $q$ were ignored.
Consequently, the threshold of the amplitude was given in Eq.~\eqref{eq:dthHYKN} and the abundance~\eqref{eq:betaana} could be approximated by
\begin{align}
    \beta_{0}\sim \exp\qty(-0.005q^4\sigma_{\textmd{H}}^{-2}),
    \end{align}
which was much more suppressed than~\eqref{eq:betasup} for $\sigma_{\textmd{H}}\ll1$.

\section{Spin distribution function}
\label{E_AM}

Since we have defined RMS $\bar{a}_{*}$ as a function of $\nu$, 
we can obtain the probability distribution for the spin $P(\bar{a}_{*})$ by solving $\bar{a}_{*}(\nu)=\bar{a}_{*}$ for $\nu$ 
and substituting it into the number density for the density peak, $\mathcal{N}_{\textmd{pk},\nu}(\nu)$, as
\begin{align}
  P(\bar{a}_{*})d\bar{a}_{*}\propto 
  \mathcal{N}_{\textmd{pk},\nu}(\nu(\bar{a}_{*}))\left|\frac{d\nu}{d\bar{a}_{*}}\right|d\bar{a}_{*} 
  \end{align}
  up to the normalization constant. 
See Eq.~\eqref{eq:Nn} for the expression and interpretation of $\mathcal{N}_{\textmd{pk},\nu}(\nu)$.

Here we have a few points to note about the distribution function $P(\bar a_*)$. 
In the derivation of $\bar a_*$, we have taken ensemble averages over $w_{i}$ and $\lambda_{i}$, with $\nu$ being fixed.
Therefore, this RMS $\bar{a}_{*}$ should be interpreted as the typical value of spin averaged over the second spatial derivatives $\lambda_i$ of the density fluctuation around the peak with a fixed amplitude $\nu$.
Without the average, the distribution function should depend on the second spatial derivatives.
Namely, in the probability distribution $P(\bar a_*)$,
the dependence on the variance does not appear explicitly.
Additionally, for the evaluation of $P(\bar a_*)$, we neglect the influence of ellipticity on the PBH formation criterion.
If the condition of ellipticity were taken into account, cases with large ellipticity might be selectively eliminated
\footnote{
  According to our formulation, the formal expression for the probability distribution of the dimensionless spin parameter $a_*$ can be given by 
  \begin{equation}
    P(a_*)da_*\propto \left[\int  \mathcal{N}_{\textmd{pk},9} \qty(\nu,\vec{\lambda},\vec{w},\tilde{\mathcal{D}}_{B},\tilde{\mathcal{D}}_{C})
    d\vec{\lambda}d\vec{w}d\tilde{\mathcal{D}}_{B}d\tilde{\mathcal{D}}_{C} 
    \theta\qty[1-h(a_*, \vec{w},\tilde{\mathcal{D}}_{B},\tilde{\mathcal{D}}_{C})] 
    \frac{\partial\qty(\nu,\vec{\lambda},\vec{w},\tilde{\mathcal{D}}_{B},\tilde{\mathcal{D}}_{C})}{\partial \qty(a_*,\vec{\lambda},\vec{w},\tilde{\mathcal{D}}_{B},\tilde{\mathcal{D}}_{C})}
    \right]da_* 
  \end{equation}
  up to the normalization constant. 
  See Eq.~\eqref{eq:PQ3} for the definition of $\mathcal{N}_{\textmd{pk},9}\qty(\nu,\vec{\lambda},\vec{w},\tilde{\mathcal{D}}_{B},\tilde{\mathcal{D}}_{C})$.
We have also used the fact that the ellipticity $(\tilde{\alpha},\tilde{\beta},\tilde{\gamma})$ can be expressed in terms of $(a_*, \vec{w},\tilde{\mathcal{D}}_{B},\tilde{\mathcal{D}}_{C})$, 
because the ellipticity parameters can be expressed in terms of $\qty(\vec{w},\tilde{\mathcal{D}}_{A}, \tilde{\mathcal{D}}_{B},\tilde{\mathcal{D}}_{C})$
and the distribution of $\tilde{\mathcal{D}}_{A}$ is identical to that of $\nu$, which can be replaced with $a_*$ 
through the expression $a_*^2=a_*^2\qty(\nu,\vec{\lambda},\vec{w},\tilde{\mathcal{D}}_{B},\tilde{\mathcal{D}}_{C})$ implicitly given by \eqref{eq:as2}. 
The integration with respect to the parameters $\vec{\lambda}$, $\vec{w}$, $\tilde{\mathcal{D}}_{B}$ and $\tilde{\mathcal{D}}_{C}$ is 
quite cumbersome because the integrand is dependent on all parameters. 
Thus we evaluate the probability distribution of the mean value $\bar a_*$ instead of performing the octuple integral.}, and the distribution in large spin regions could be further suppressed.
Nevertheless, we may expect that $P(\bar a_*)$ helps understand the qualitative behavior of the spin distribution.

\begin{figure}[htbp]
    \begin{center}
     \includegraphics[clip,width=8cm]{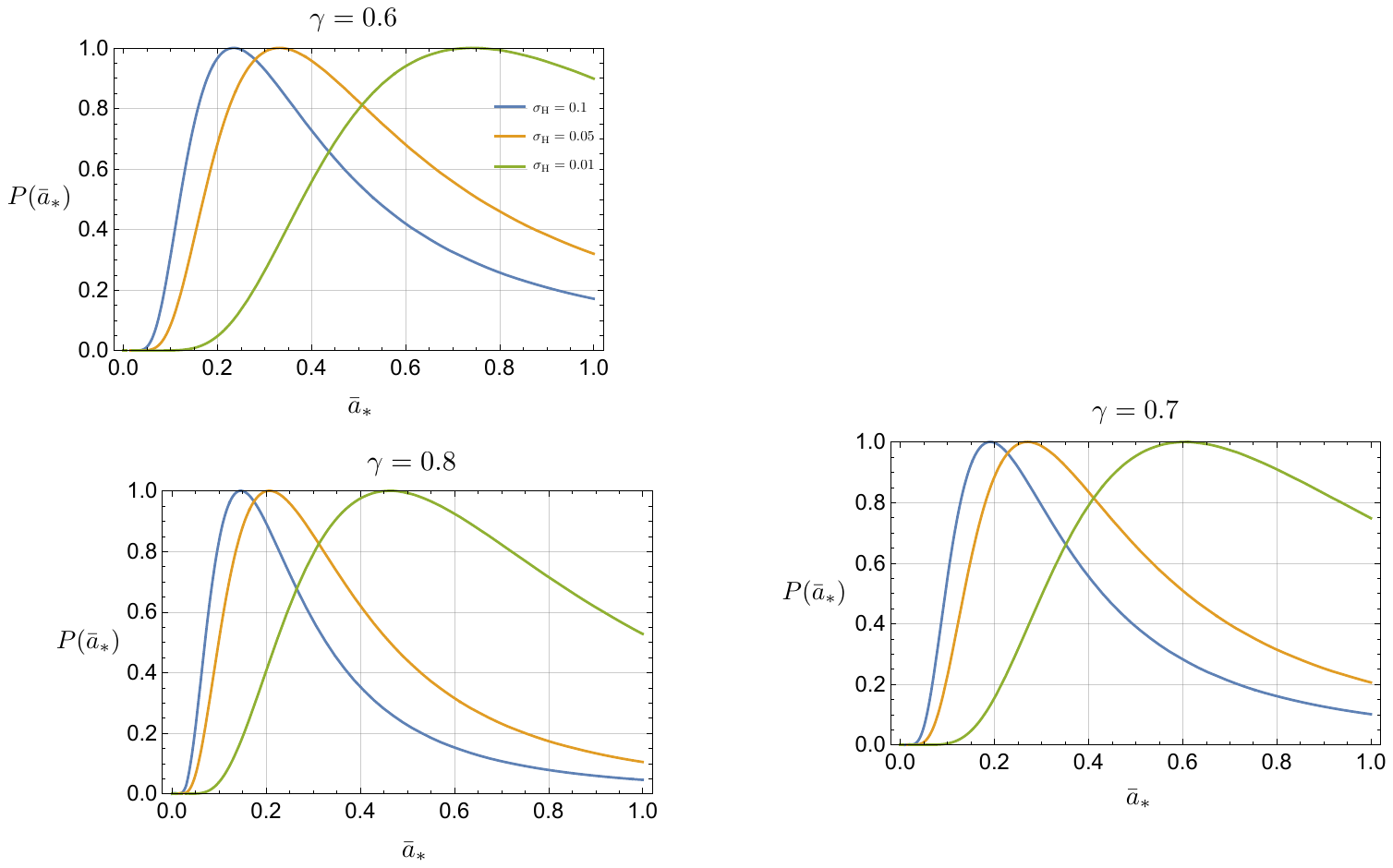} 
     \includegraphics[clip,width=8cm]{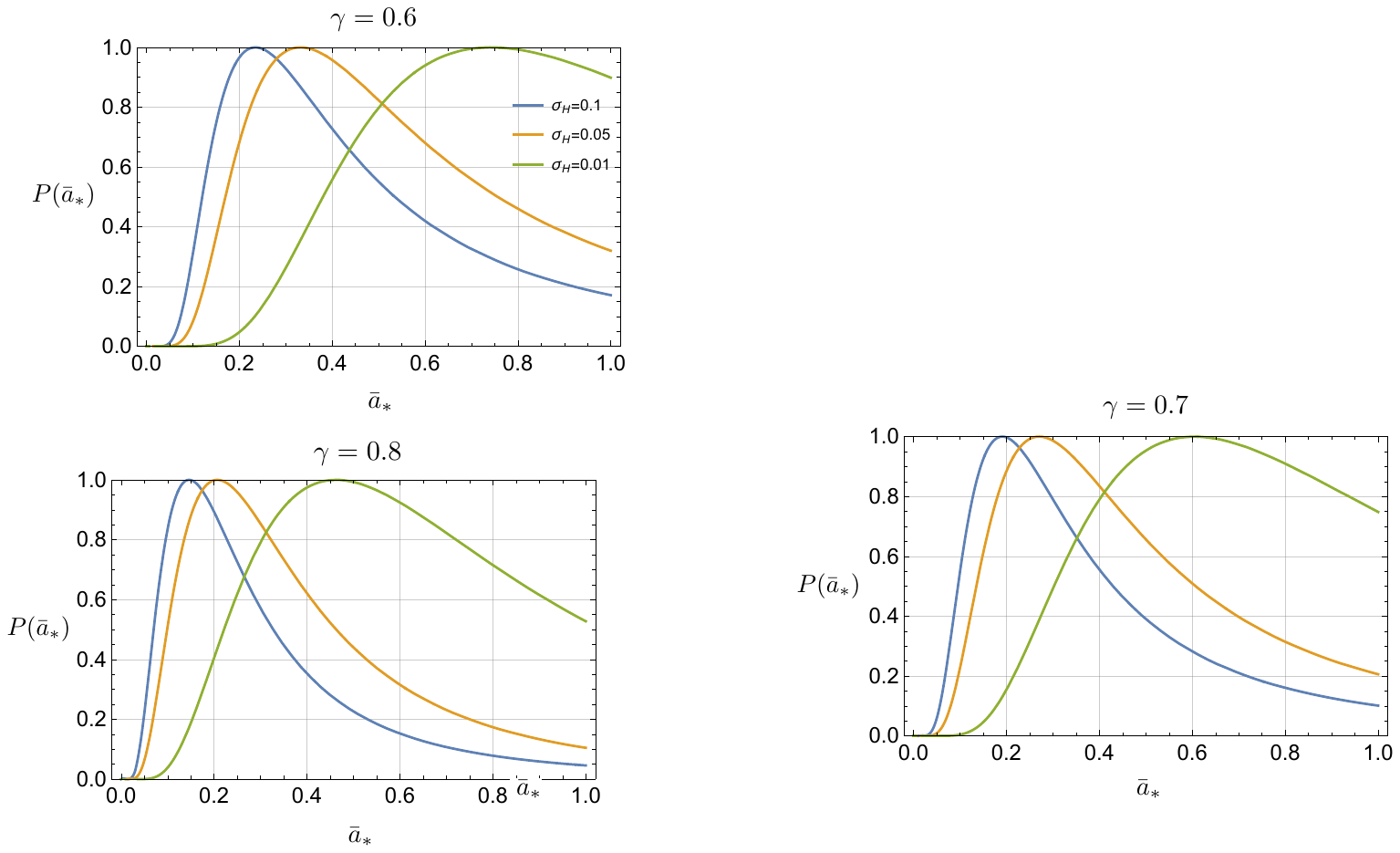} 
     \includegraphics[clip,width=8cm]{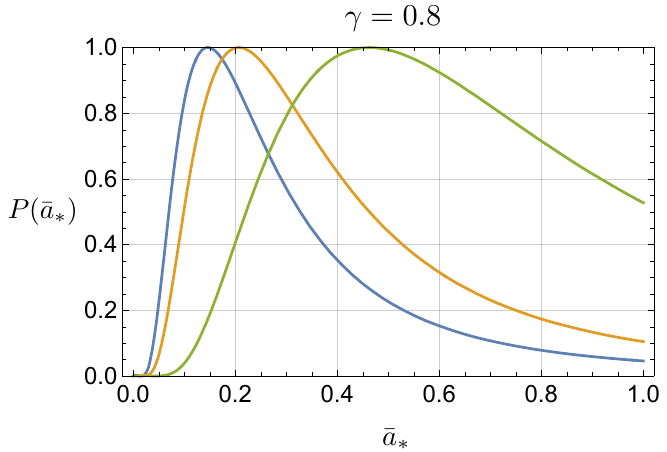} 
     \caption{The probability distribution of spin $P(\bar{a}_{*})$ for $\gamma=0.6$ (upper left), $\gamma=0.7$ (upper right) and $\gamma=0.8$ (lower).
     Each graph is normalized by its maximum value.
     With $\sigma_{\textmd{H}}=0.1$ and $\gamma=0.8$, $P(\bar{a}_{*})$ peaks at $\bar{a}_{*}<1$, while the other cases have peak at $\bar{a}_{*}=1$.}
     \label{fig:Pa}
  \end{center}
   \end{figure}

The probability distribution of spin $P(\bar{a}_{*})$ is shown in Fig.~\ref{fig:Pa}.
We can see that, for $\gamma=0.8$, the distribution has a maximum at the subcritical value of spin.
For instance, $P(\bar{a}_{*})$ takes maximum at $\bar{a}_{*}=0.15$  and $0.45$ for $\sigma_{\textmd{H}}=0.1$ and $0.01$, respectively. 
From Eq.~\eqref{eq:AMEnst}, we can see that the value of $\bar a_*$ is a monotonic decreasing function of $\nu$. 
Therefore the existence of the peak in the distribution function $P(a_*)$ originates from the functional form of 
$\mathcal{N}_{\textmd{pk},\nu}(\nu)$.  
Since the value of $\bar a_*$ is a monotonic decreasing function of $\nu$, the position of the maximum of $P(\bar{a}_{*})$ decreases when the maximum position of $\mathcal{N}_{\textmd{pk},\nu}$ is larger.
As we show in Fig.~\ref{fig:Pn}, the position of the maximum of $\mathcal{N}_{\textmd{pk},\nu}(\nu)$, $\nu_{\textmd{max}}$, increases with $\gamma$.
Thus, for a narrow power spectrum, $P(\bar{a}_{*})$ takes a maximum at a smaller value of $\bar{a}_{*}$.
\footnote{In Ref.~\cite{Harada:2017fjm}, from the expression \eqref{eq:ashykn}, the authors assumed that 
$\left(\bar a_*^{\rm HYKN}\right)^{-2}$ obey a Gaussian distribution. Even with this treatment, we may have 
a maximum in $\bar a_*^{\rm HYKN}<1$ if we take $q$ smaller.
}

\section{Joint probability}
\label{Pam}

Let us derive a joint probability distribution of non-dimensional mass and spin induced by the overdensity.
We shall define the dimensionless mass parameter as
\begin{align}
    m:=\frac{M}{M_{*}}=\sqrt{8}\qty(\frac{\nu}{\gamma x})^{3/2}\frac{1}{\sqrt{B(e, p)}} \label{eq:m},
\end{align}
where
\begin{align}
    M_{*}:=\frac{4\pi}{3}\eta_{0}R^3_{*}
  \end{align}
is the reference mass.
$R_{*}=\sqrt{3}\sigma_{1}/\sigma_{2}$ is the parameter that characterizes the length scale of fluctuation.
Although we do not specify the explicit value of $M_{*}$, it can be obtained once one fixes 
the power spectrum and the energy density in the early MD era.

By solving Eqs.~\eqref{eq:a} and \eqref{eq:m} with $t=t_{\rm ta}$ for $\nu$ and $x$ as $\nu=\nu(a_{\lambda}, m, e, p)$, $x=x(a_{\lambda}, m, e, p)$ and integrating over $e$ and $p$, we can obtain the joint probability as
\begin{align}
    P(a_{\lambda},m)da_{\lambda}dm=\qty[\qty(\int_{0}^{1/4} de\int_{-e}^{e} dp + \int_{1/2}^{1/4} de\int_{3e-1}^{e} dp) \mathcal{N}_{\textmd{pk},4}(\nu, x,e,p)\left|\frac{\partial(a_{\lambda}, m)}{\partial(\nu, x)}\right|^{-1}]da_{\lambda}dm,
  \end{align}
where $\mathcal{N}_{\textmd{pk},4}(\nu, x,e,p)$ is a comoving number density for the density peak with $(\nu, x,e,p)$ (See App.~\ref{PT} for detail).
Here, we have restricted the range of the integration 
so that the condition $\lambda_{1}\geq\lambda_{2}\geq\lambda_{3}\geq0$ 
will be satisfied.

\begin{figure}[htbp]
    \begin{center}
     \includegraphics[clip,width=8cm]{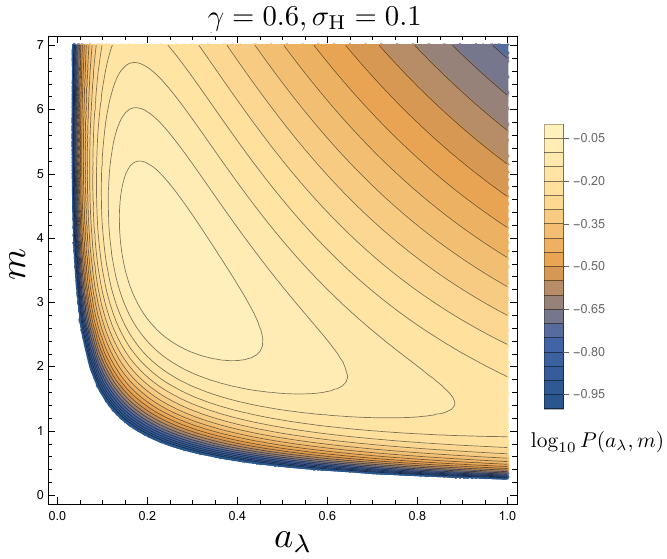} 
     \includegraphics[clip,width=8cm]{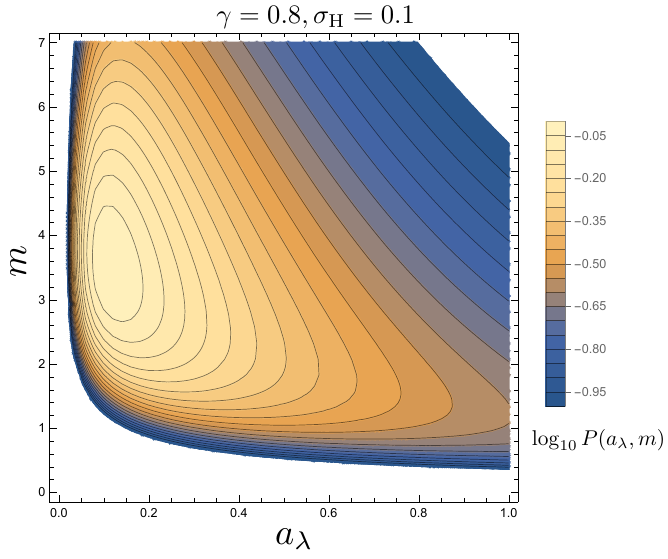} 
     \includegraphics[clip,width=8cm]{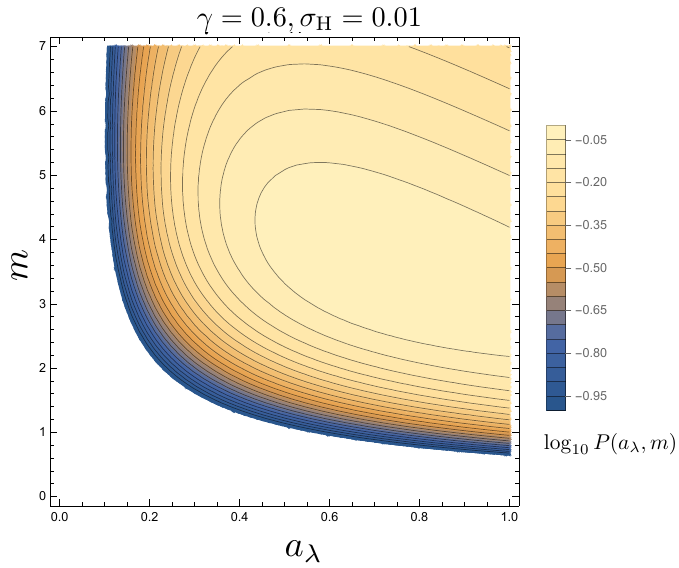} 
     \includegraphics[clip,width=8cm]{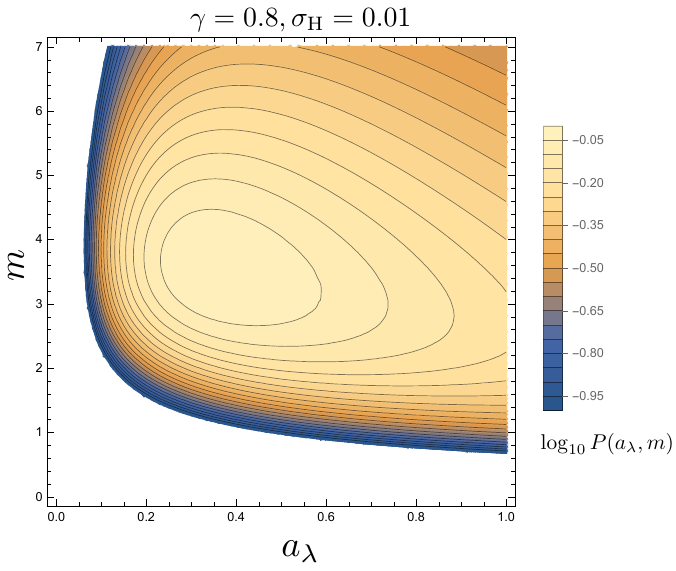} 
     \caption{The contour of the probability distribution of spin and mass $\log_{10}P(a_{\lambda}, m)$ with $\sigma_{\textmd{H}}=0.1 ,\gamma=0.6$ (upper left), $\sigma_{\textmd{H}}=0.1 ,\gamma=0.8$ (upper right), $\sigma_{\textmd{H}}=0.01 ,\gamma=0.6$ (lower left) and $\sigma_{\textmd{H}}=0.01 ,\gamma=0.8$ (lower right).
     Each graph is normalized by its maximum value.}
     \label{fig:Pam}
  \end{center}
   \end{figure}

The contours of the joint probability distribution for mass and spin are plotted in Fig.~\ref{fig:Pam}.
We have restricted the range of the spin to $a_{\lambda} \leq 1$ in our analysis, focusing on estimating the PBH spin, 
and the joint probability is normalized by its maximum value in that domain.
We observe that with $\sigma_{\textmd{H}}=0.1$, the maximum values of the joint distribution are realized around $a_{\lambda}=0.25$ and $a_{\lambda}=0.1$ for $\gamma=0.6$ and $\gamma=0.8$, respectively, consistently with the findings in the previous section.
The distribution around the maximum sharpens with increasing values of $\sigma_{\rm H}$ or $\gamma$.
The tendency also agrees with the result for $P(\bar{a}_{*})$ in the previous section.
Concerning the mass, we do not observe significant dependence of the position of the maximum on $\sigma_{\rm H}$ or $\gamma$.
In all cases shown in Fig.~\ref{Pam}, $P(a_{\lambda},m)$ takes a maximum around $m=3.5$.
However, the sharpness of the distribution around this maximum increases with larger $\sigma_{\rm H}$ or $\gamma$, similar to the result for the distribution of $a_{*}$.
Furthermore, we note that when $\sigma_{\rm H}$ and $\gamma$ are fixed, the probability distribution decreases relatively mildly when either $a_{\lambda}$ or $m$ is held constant while the other increases from the peak position.
In contrast, the slope is steeper when both $a_{\lambda}$  and $m$ increase simultaneously.
Thus, we anticipate that the probability of PBHs having both large mass and spin simultaneously is suppressed.
This tendency is more pronounced for a narrower power spectrum and a larger variance of the fluctuation.

\section{Summary and discussion}
\label{Sum}  

In this paper, we have evaluated the linear-order effect of the spin generation for primordial black holes (PBHs) formed in a matter-dominated (MD) universe.   
We followed the method in Ref.~\cite{Catelan:1996hv}.
Specifically, we employed the Zel'dovich approximation and assumed that the perturbative variables can be treated as random Gaussian fields in the linear order. 
Then we applied the peak theory of random Gaussian variables~\cite{Bardeen:1985tr,Heavens:1988}.

We calculated the first-order contribution of the perturbation on the non-dimensional Kerr parameter inside a collapsing region around a peak of density perturbation.
Then we evaluated the root mean square (RMS) $\bar{a}_{*}$ of the spin, which is defined as an ensemble average over the second spatial derivatives of the density perturbation and that of the gravitational potential, at the turn-around time $t_{\textmd{ta}}$.
We found that the RMS decreases with $\nu:=\delta_{\textmd{pk}}/\sigma_{\rm H}$, where $\delta_{\textmd{pk}}$ is the amplitude of the fluctuation and $\sigma_{\textmd{H}}$ is the root-mean-square deviation of the density fluctuation at the horizon reentry. 
Additionally, we also found $\bar{a}_{*}(t_{\textmd{ta}})\propto \sigma^{-1/2}_{\textmd{H}}$
with a fixed value of $\nu$.
This behavior can be attributed to the fact that smaller initial fluctuations require a longer time to the turn-around and the growth of their spin can last for a longer time until the turn-around.

We derived the PBH formation threshold for the density fluctuation at the horizon reentry, denoted as $\delta_{\textmd{H,th}}$, based on the non-extremal condition $\bar{a}_{*}(\delta_{\textmd{H,th}})\leq1$.
In contrast with the PBH formation in a radiation-dominated (RD) universe, the obtained threshold depends on $\sigma_{\textmd{H}}$ and increases monotonically with $\sigma_{\textmd{H}}$, and takes $\delta_{\textmd{H,th}}\sim 0.1$ for $\sigma_{\textmd{H}}=10^{-1}$.

By combining the threshold derived from the hoop conjecture~\cite{Thorne:1972ji,Harada:2016mhb}, we examined the impact of angular momentum and anisotropic collapse on the production rate of PBHs.
We found that, with $\gamma=0.6$ and $0.8$, the suppression on the PBH abundance due to the spin effect is stronger compared to 
that of the anisotropic collapse for 
$\sigma_{\textmd{H}}\lesssim 10^{-3}$ and $\sigma_{\textmd{H}}\lesssim 4\times 10^{-4}$, respectively. 
Therefore, we conclude that the effect of the threshold obtained from the angular momentum on the production rate is effective when the variance of the fluctuation at the horizon reentry is sufficiently small, and this tendency weakly depends on the value of $\gamma$.
We also compared these results with those in HYKN~\cite{Harada:2017fjm}, where correlations of independent variables associated with the peak theory are not taken into account, and the turn-around time is determined independently of $\nu$.
Consequently, we found that the suppression of the PBH abundance is considerably weakened in our improved estimation.

By using the expression of $\bar{a}_{*}$ as a function of $\nu$, we transformed the probability density of $\nu$ given by $\mathcal{N}_{\textmd{pk},\nu}(\nu)$ into the probability density of $\bar{a}_{*}$, $P(\bar{a}_{*})$. 
Then we found that $P(\bar{a}_{*})$ typically has a maximum in $\bar{a}_{*}=O(10^{-1})$ for $\sigma_{\rm H}=O(10^{-2})-O(10^{-1})$, depending on the value of $\gamma$, which characterizes the width of the power spectrum.
This result implies that, in an MD universe, PBHs can have a larger spin value than the typical value $O(10^{-3})$ in an RD universe~\cite{Harada:2020pzb}.
We also computed the joint probability distribution for the dimensionless mass and spin of PBH, denoted as $P(a_{\lambda},m)$.
We also found that the distribution around the peak position of $P(a_{\lambda},m)$ gets sharper for larger values of $\sigma_{\rm H}$ and $\gamma$.

While we have investigated the probability distribution of the PBH spin in an MD era and its impact on the PBH abundance, there are still aspects that require further exploration.
We have compared the effect of spin on PBH production with that of anisotropic collapse.
In discussing the effect of anisotropic collapse, we assumed the configuration overdensity at the horizon reentry to be spherical, neglecting the effect of initial non-sphericity.
We have also neglected the effect of ellipticity on the PBH formation on the evaluation of the probability distribution of the spin.
Furthermore, other mechanisms such as inhomogeneity~\cite{Kokubu:2018fxy} and velocity dispersion~\cite{Harada:2022xjp} could affect PBH formation in an MD universe.
A comprehensive comparison of spin effects with other factors on the production rate of PBHs also awaits further investigation.

\section*{Acknowledgements}

This work was partially supported by JSPS KAKENHI Grant Numbers 24KJ1223 (DS), 24K07027 (TH, CY), 20H05853 (TH, CY), 21K20367 (YK), 23KK0048 (YK), and 20H05850 (CY).

\appendix

\section{Peak theory for random Gaussian fields}
\label{PT}

In this appendix, we review the peak theory for random Gaussian fields~\cite{Bardeen:1985tr,Heavens:1988}.
We shall consider the following quantities as random variables following a Gaussian distribution.
\begin{align}
&V_{i}:=\qty{\delta,\delta_{i},\delta_{ij},\mathcal{\tilde{D}}_{ij}}. 
    \end{align}
Here, we have defined
\begin{align}
    &\delta_{i}:=\frac{\partial\delta}{\partial q^{i}}, \\
    &\delta_{ij}:=\frac{\partial^2\delta}{\partial q^{i}\partial q^{j}}, \\
    &\mathcal{\tilde{D}}_{ij}:=\frac{\mathcal{D}_{ij}}{4\pi\rho_{\textmd{b}}a^2}=\int\frac{d^3\vec{k}}{(2\pi)^3}\frac{k_{i}k_{j}}{k^2}\delta_{\vec{k}}e^{i\vec{k}\cdot\vec{q}}.
\end{align}
The joint probability for these 16 variables is written as
\begin{align}
    f(V_{i})d^{16}V_{i}=\frac{1}{(2\pi)^8\sqrt{\det M}}\exp\qty[-\frac{1}{2}\qty(V_{i}-\bar{V}_{i})M_{ij}^{-1}\qty(V_{j}-\bar{V}_{j})]d^{16}V_{i},
\end{align}
where $M_{ij}:=\langle V_{i}V_{j} \rangle$ is the correlation function and $\bar{V}_{i}$ denotes the mean value of $V_{i}$.
The non-zero correlations are given as
\begin{align}
&\left\langle\delta^2\right\rangle=3\left\langle\delta \tilde{\mathcal{D}}_{11}\right\rangle=5\left\langle \tilde{\mathcal{D}}^2_{11}\right\rangle=15\left\langle \tilde{\mathcal{D}}_{11}\tilde{\mathcal{D}}_{22}\right\rangle=\cdots=\sigma^2_{0}, \\
    &\left\langle\delta\delta_{11}\right\rangle=–\left\langle\delta_{1}\delta_{1}\right\rangle=\frac{5}{3}\left\langle\delta_{11}\tilde{\mathcal{D}}_{11}\right\rangle=5\left\langle\delta_{12}\tilde{\mathcal{D}}_{12}\right\rangle=\cdots=-\frac{\sigma^2_{1}}{3}, \\    &\langle\delta^2_{11}\rangle=3\langle\delta_{11}\delta_{22}\rangle=3\langle\delta^2_{12}\rangle=\cdots=\frac{\sigma^2_{2}}{5}.
\end{align}

Changing the variables as
\begin{align}
    &x:=-\frac{1}{\sigma_{2}}(\delta_{11}+\delta_{22}+\delta_{33}), \quad y:=-\frac{1}{2\sigma_{2}}(\delta_{11}-\delta_{33}), \quad z:=-\frac{1}{2\sigma_{2}}(\delta_{11}-2\delta_{22}+\delta_{33}), \\
    &\nu=\frac{\delta}{\sigma_{0}}, \quad \tilde{\delta}_{i}:=\frac{\delta_{i}}{\sigma_{1}} \quad (i=1,2,3), \quad \tilde{\delta}_{ij}:=\frac{\delta_{ij}}{\sigma_{2}} \quad (i\neq j),\\
    &\tilde{\mathcal{D}}_{A}:=\frac{1}{\sigma_{0}}\qty(\tilde{\mathcal{D}}_{11}+\tilde{\mathcal{D}}_{22}+\tilde{\mathcal{D}}_{33}), \quad \tilde{\mathcal{D}}_{B}:=\frac{1}{2\sigma_{0}}\qty(\tilde{\mathcal{D}}_{11}-\tilde{\mathcal{D}}_{33}), \quad \tilde{\mathcal{D}}_{C}:=\frac{1}{2\sigma_{0}}\qty(\tilde{\mathcal{D}}_{11}-2\tilde{\mathcal{D}}_{22}+\tilde{\mathcal{D}}_{33}), \\
    &w_{1}:=-\frac{\tilde{\mathcal{D}}_{23}}{\sigma_{0}}, \quad w_{2}:=-\frac{\tilde{\mathcal{D}}_{31}}{\sigma_{0}}, \quad w_{3}:=-\frac{\tilde{\mathcal{D}}_{12}}{\sigma_{0}},
\end{align}
we can find that $\left\langle\tilde{\mathcal{D}}_{A}\nu\right\rangle=1$, which means tnat $\tilde{\mathcal{D}}_{A}$ is perfectly correlated with $\nu$ and thus, we drop $\tilde{\mathcal{D}}_{A}$.
In the following, we shall diagonalize $\delta_{ij}$ and denote their eigenvalues as $\lambda_{i}$.

According to Refs.~\cite{Bardeen:1985tr,Heavens:1988}, by imposing the conditions for the peak, $\delta_{i}=0$ and $\lambda_{i}>0$, the comoving number density of the density peak with $\qty(\nu,\vec{\lambda},\vec{w},\tilde{\mathcal{D}}_{B},\tilde{\mathcal{D}}_{C})$ being within infinitesimal ranges can be obtained as
\begin{align}
    &\mathcal{N}_{\textmd{pk},9}\qty(\nu,\vec{\lambda},\vec{w},\tilde{\mathcal{D}}_{B},\tilde{\mathcal{D}}_{C})d\nu d\vec{\lambda}d\vec{w}d\tilde{\mathcal{D}}_{B}d\tilde{\mathcal{D}}_{C}=\bar{A}\qty(\frac{\sigma_{2}}{\sigma_{1}})^3\exp(-Q_{3})\lambda_{1}\lambda_{2}\lambda_{3}(\lambda_{2}-\lambda_{3})(\lambda_{1}-\lambda_{3})(\lambda_{1}-\lambda_{2})d\nu d\vec{\lambda}d\vec{w}d\tilde{\mathcal{D}}_{B}d\tilde{\mathcal{D}}_{C}, \label{eq:PQ3} \\
    &\bar{A}:=\frac{3^{11/2}5^5}{2^{13/2}\pi^{11/2}(1-\gamma^2)^3}, \\
    &2Q_{3}:=\nu^2+\frac{(x-\nu\gamma)^2}{1-\gamma^2}+15y^2+\frac{5\qty(\tilde{D}_{B}-y\gamma)^2}{1-\gamma^2}+5z^2+\frac{15\qty(\tilde{D}_{C}-z\gamma)^2}{1-\gamma^2}+15\frac{w_{1}^2+w_{2}^2+w_{3}^2}{1-\gamma^2}.
    \end{align}
In the following, we will use the notation $\qty(\nu,\vec{\lambda},\vec{w},\tilde{\mathcal{D}}_{B},\tilde{\mathcal{D}}_{C})$ to denote the values at the peaks.
    
As long as we focus on evaluating the spins without considering the conditions of ellipticity of the fluctuation, the statistical properties of $\tilde{\mathcal{D}}_{B}$ and $\tilde{\mathcal{D}}_{C}$ can be ignored because the spins are independent of them.
By integrating $\tilde{\mathcal{D}}_{B}$ and $\tilde{\mathcal{D}}_{C}$, we obtain
\begin{align}
    &\mathcal{N}_{\textmd{pk}}\qty(\nu,\vec{\lambda},\vec{w})d\nu d\vec{\lambda}d\vec{w}=\frac{\bar{B}}{R^3_{*}}\exp(-Q_{4})F(\lambda)d\nu d\vec{\lambda}d\vec{w}, \label{eq:PQ4} \\
    &R_{*}:=\sqrt{3}\frac{\sigma_{1}}{\sigma_{2}}, \quad \bar{B}:=\frac{3^{9/2}5^4}{2^{11/2}\pi^{9/2}(1-\gamma^2)^2}, \quad F(\lambda):=\frac{27}{2}\lambda_{1}\lambda_{2}\lambda_{3}(\lambda_{2}-\lambda_{3})(\lambda_{1}-\lambda_{3})(\lambda_{1}-\lambda_{2}), \\
    &2Q_{4}:=\nu^2+\frac{(x-\nu\gamma)^2}{1-\gamma^2}+15y^2+5z^2+15\frac{w_{1}^2+w_{2}^2+w_{3}^2}{1-\gamma^2}.
    \end{align}
By changing the variables as $\vec{\lambda}\rightarrow \vec{\tilde{\lambda}}:=(x,p,e)$, which are given by
\begin{align}
x=\lambda_{1}+\lambda_{2}+\lambda_{3}, \quad e=\frac{\lambda_{1}-\lambda_{3}}{2x}, \quad p=\frac{\lambda_{1}-2\lambda_{2}+\lambda_{3}}{2x},
\end{align}
we obtain
\begin{align}   \mathcal{N}_{\textmd{pk}}\qty(\nu,\vec{\tilde{\lambda}},\vec{w})d\nu d\vec{\tilde{\lambda}}d\vec{w}&=\mathcal{N}_{\textmd{pk},w}(\vec{w})d\vec{w}\times \mathcal{N}_{\textmd{pk},4}(\nu, x,e,p) d\nu d\vec{\tilde{\lambda}}.
    \label{eq:Pw} 
        \end{align}
Here, we have defined
        \begin{align}   
    &\mathcal{N}_{\textmd{pk},w}(\vec{w})d\vec{w}:=\qty(\frac{15}{2\pi(1-\gamma^2)})^{3/2}\exp(-Q_{w})d\vec{w}, \\
    &\mathcal{N}_{\textmd{pk},4}(\nu, x,e,p) d\nu d\vec{\tilde{\lambda}}:=\frac{3^{2}5^{5/2}}{(2\pi)^3(1-\gamma^2)^{1/2}}\frac{x^8}{R^3_{*}}\exp(-\tilde{Q}_{4})W(e,p)d\nu d\vec{\tilde{\lambda}}, \label{eq:N4}
    \\
    2Q_{w}:=15\frac{w_{1}^2+w_{2}^2+w_{3}^2}{1-\gamma^2}, \quad &2\tilde{Q}_{4}:=\nu^2+\frac{(x-\nu\gamma)^2}{1-\gamma^2}+5x^2(3e^2+p^2), \quad W(e,p)=e(e^2-p^2)B(e,p).
    \end{align}

From this expression, we see that the distribution of $\vec{w}$ is independent of that of $\nu$ and $\tilde{\lambda}$.
Then, we can obtain~\cite{Catelan:1996hv}
\begin{align}
    \left\langle \mathcal{D}_{ij}\mathcal{D}_{kl} \right\rangle_{w_{i}}&=\frac{1-\gamma^2}{15}(4\pi\rho_{\textmd{b}}a^2\sigma_{0})^2\qty(\delta_{ik}\delta_{jl}+\delta_{il}\delta_{jk}+\delta_{ij}\delta_{kl}).\label{eq:DDA}
  \end{align}

We can determine the number density with height $\nu$ by integrating $\mathcal{N}_{\textmd{pk}}(\nu, x,e,p)$ over $0\leq x\leq\infty$, and integrating over $p$ and $e$ within the following range:
\begin{align}
&0\leq e\leq\frac{1}{4} \; \text{and} \; -e\leq p\leq e, \quad \frac{1}{4}\leq e\leq\frac{1}{2} \; \text{and} \; 3e-1\leq p\leq e. \label{eq:Integ}
    \end{align}
Here, the restriction of the integration domain of $x$ corresponds to the requirement that the peak should be a local maximum: $\lambda_{3}\geq0$.
After the integration, we obtain the differential density of peaks in the range of $\nu$ and $\nu+\delta\nu$
\begin{align}
    &\mathcal{N}_{\textmd{pk},\nu}(\nu)d\nu\propto \frac{1}{R^3_{*}}e^{-\nu^2/2}\int_{0}^{\infty}dxf(x)\frac{1}{\qty[2\pi(1-\gamma^2)]^{1/2}}\exp\qty[-\frac{(x-\gamma\nu)^2}{2(1-\gamma^2)}] d\nu, \label{eq:Nn} \\ 
    &f(x):=\frac{(x^3-3x)}{2}\qty{\textmd{erf}\qty[\qty(\frac{5}{2})^{1/2}x]+\textmd{erf}\qty[\qty(\frac{5}{2})^{1/2}\frac{x}{2}]}+\qty(\frac{2}{5\pi})^{1/2}\qty[\qty(\frac{31x^2}{4}+\frac{8}{5})e^{-5x^2/8}+\qty(\frac{x^2}{2}-\frac{8}{5})e^{-5x^2/2}],
  \end{align}
and the distribution for the conditional probability for $\tilde{\lambda}$ with fixed $\nu$ can be defined as
\begin{align}
    P(x,e,p|\nu):=\frac{\mathcal{N}_{\textmd{pk},4}(\nu, x,e,p)}{\mathcal{N}_{\textmd{pk},\nu}(\nu)}. \label{eq:Pxepn}
  \end{align}
  
\begin{figure}[htbp]
    \begin{center}
     \includegraphics[clip,width=10cm]{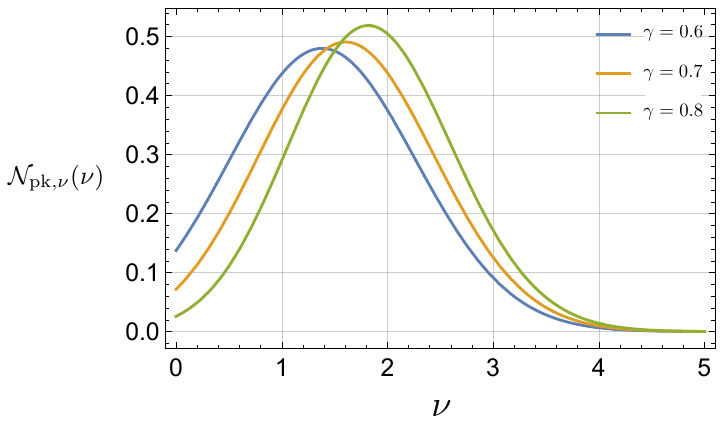} 
     \caption{The probability distribution of the number density of the peak $\mathcal{N}_{\textmd{pk},\nu}(\nu)$ for $\gamma=0.6$ (blue), $0.7$ (orange) and $\gamma=0.8$ (green).}
     \label{fig:Pn}
  \end{center}
   \end{figure}

In Fig.~\ref{fig:Pn}, $\mathcal{N}_{\textmd{pk},\nu}(\nu)$ for several values of $\gamma$ is shown.
It is observed that $\mathcal{N}_{\textmd{pk},\nu}(\nu)$ has a maximum at some value $\nu_{\textmd{max}}>0$, which reflects the fact that there is a correlation between $\nu$ and $x$: $\langle\nu x\rangle=\gamma$.
From Fig.~\ref{fig:Pn}, we can see that $\nu_{\textmd{max}}$ and the value of $\mathcal{N}_{\textmd{pk},\nu}({\nu_{\textmd{max}})}$ increase as $\gamma$ increases.

It is worth noting that, since the conditions for density peaks ($\delta_{i}=0$ and $\lambda_{i}>0$) were imposed in its derivation (See Ref.~\cite{Bardeen:1985tr,Heavens:1988} for detail),  $\mathcal{N}_{\textmd{pk}}\qty(\nu,\vec{\tilde{\lambda}},\vec{w})$ can be interpreted as a probability distribution function for peaks that may collapse into PBHs, up to normalization.
Therefore, in Sec.~\ref{E_AM} and Sec.~\ref{Pam}, we compute the PBH spin distribution and the joint distribution for PBH mass and spin using $\mathcal{N}_{\textmd{pk},4}(\nu, x,e,p)$ and $\mathcal{N}_{\textmd{pk},\nu}(\nu)$, respectively.

  \section{Turn-around time}
  \label{TA}

In this section, we derive the turn-around time for a spherical overdensity based on the discussion in Ref.~\cite{Harada:2022xjp}. 
First, let us introduce a closed FLRW universe filled with dust fluid as a model for the overdensity.
The Friedmann equation can be written as
\begin{align}
    H^2(t)=H^2(t_{\textmd{H}})\qty[\Omega_{m}(t_{\textmd{H}})\qty(\frac{a(t_{\textmd{H}})}{a(t)})^3+(1-\Omega_{m}(t_{\textmd{H}}))\qty(\frac{a(t_{\textmd{H}})}{a(t)})^2], \label{eq:Closed}
  \end{align}
where we have chosen the horizon reentry time as a reference time and $\Omega_{m}:=\frac{8\pi\rho}{3H^2}$ is the density parameter for the dust fluid.
The spatial curvature can be defined by
\begin{align}
    K:=(\Omega_{m}(t_{\textmd{H}})-1)(aH(t_{\textmd{H}}))^2>0.
  \end{align}
Eq.~\eqref{eq:Closed} can be solved to give a parameterized solution
\begin{align}
    &\frac{a(t)}{a(t_{\textmd{H}})}=\frac{\Omega_{m}(t_{\textmd{H}})}{2 (\Omega_{m}(t_{\textmd{H}})-1)}(1-\cos\theta) , \\
    &t(\theta)=\frac{\Omega_{m}(t_{\textmd{H}})}{2H(t_{\textmd{H}})(\Omega_{m}(t_{\textmd{H}})-1)^{3/2}}(\theta-\sin\theta), \label{eq:solt}
  \end{align}
where the parameter $\theta$ is related to the conformal time $\tau$ 
with
$\theta=\sqrt{K}\tau$.

By using Eq.~\eqref{eq:FRLW1} with $K=0$ and
\begin{align}
    &\rho=\frac{3}{8\pi}H^2\Omega_{m},
  \end{align}
we can see that the density fluctuation at the horizon reentry satisfies
\begin{align}
    &\qty(\delta(t_{\textmd{H}})+1)\qty(\frac{H(t_{\textmd{H}})}{H_{\textmd{b}}(t_{\textmd{H}})})^2=\Omega_{m}(t_{\textmd{H}}).
  \end{align}
Note that this relation is valid for any choice of gauge.
In the synchronous comoving gauge, the density fluctuation is given by
\begin{align}
    &\delta_{C}(t_{\textmd{H}})=\frac{3}{5}\delta_{\textmd{UH}}(t_{\textmd{H}})=\frac{3}{5}\qty(\Omega_{m}(t_{\textmd{H}})-1),
  \end{align}
where $\delta_{C}$ and $\delta_{\textmd{UH}}$ denote the density fluctuation in the comoving slicing and the uniform Hubble slicing, respectively.
Here we have used the relation between perturbations in different slicing conditions derived with the long-wavelength approximation~\cite{Harada:2015yda}.

The turnaround time can be defined as the time of the maximum expansion, denoted as $t_{\textmd{ta}}:=t(\theta=\pi)$.
Assuming that $\delta(t_{\textmd{H}})\ll1$, we can compute it as
\begin{align}
    t_{\textmd{ta}}&=\frac{\Omega_{m}(t_{\textmd{H}})}{2H(t_{\textmd{H}})(\Omega_{m}(t_{\textmd{H}})-1)^{3/2}}\pi  \nonumber \\
    &\simeq\frac{\pi}{2H(t_{\textmd{H}})}\qty(\frac{5}{3}\nu\sigma_{\textmd{H}})^{-3/2}. \label{eq:tta}
  \end{align}
Here, we have used the relation $\delta_{C}(t_{\textmd{H}})=\nu\sigma_{\textmd{H}}$.

\bibliography{bibs/hoge}
\bibliographystyle{unsrt.bst}

\end{document}